\documentclass[sigconf]{acmart}

\usepackage{booktabs}
\usepackage{algorithm}
\usepackage{algpseudocode}
\usepackage{bbold}
\usepackage{enumitem}
\usepackage{subfigure}
\usepackage{colortbl}
\usepackage{algorithm}
\usepackage{algpseudocode}
\usepackage{multirow}
\usepackage[export]{adjustbox}
\usepackage{amssymb}
\usepackage{setspace}
\usepackage{graphicx}
\usepackage{xfrac}

\newcommand{\todo}[1]{\textcolor{red}{#1}}
\newcommand{\sz}[1]{\textcolor{blue}{#1}}

\newtheoremstyle{mydef}% style name
{2ex}% above space
{2ex}% below space
{\itshape}% body font
{}% indent amount
{\scshape}% head font
{: }% post head punctuation
{0.5em}% space after theorem head
{}% head spec 
\theoremstyle{mydef}
\newtheorem{mydef}{Definition}

\setlist[itemize]{leftmargin=*}

\setcopyright{rightsretained} 

\begin{document}

\copyrightyear{2020}
\acmYear{2020}
\setcopyright{acmcopyright}
\acmConference[CIKM '20]{Proceedings of the 29th ACM International Conference on Information and Knowledge Management}{October 19--23, 2020}{Virtual Event, Ireland}
\acmBooktitle{Proceedings of the 29th ACM International Conference on Information and Knowledge Management (CIKM '20), October 19--23, 2020, Virtual Event, Ireland}
\acmPrice{15.00}
\acmDOI{10.1145/3340531.3412019}
\acmISBN{978-1-4503-6859-9/20/10}

\title{Generating Categories for Sets of Entities}
\author{Shuo Zhang}\authornote{Work done while visiting CMU, USA.}
\affiliation{%
  \institution{Bloomberg}
  \city{London}
  \country{United Kingdom}
}
\email{szhang611@bloomberg.net}
\author{Krisztian Balog}
\affiliation{%
  \institution{University of Stavanger}
  \city{Stavanger}
  \country{Norway}
}
\email{krisztian.balog@uis.no}
\author{Jamie Callan}
\affiliation{%
  \institution{Carnegie Mellon University}
  \city{Pittsburgh}
  \country{USA}
}
\email{callan@cs.cmu.edu}

\begin{abstract}
Category systems are central components of knowledge bases, as they provide a hierarchical grouping of semantically related concepts and entities.  They are a unique and valuable resource that is utilized in a broad range of information access tasks.
To aid knowledge editors in the manual process of expanding a category system, this paper presents a method of generating categories for sets of entities.
First, we employ neural abstractive summarization models to generate candidate categories.  Next, the location within the hierarchy is identified for each candidate.  Finally, structure-, content-, and hierarchy-based features are used to rank candidates to identify by the most promising ones (measured in terms of specificity, hierarchy, and importance).  We develop a test collection based on Wikipedia categories and demonstrate the effectiveness of the proposed approach.

\end{abstract}




\maketitle

\section{Introduction}
\label{sec:int}

Category systems provide a hierarchical and topical organization of entities and concepts.
They are meant to help arrange and access topically related items, and are immensely useful. 
Compared to type hierarchies of knowledge bases, such as the DBpedia ontology or Freebase types, we are focusing on category systems that are much finer-grained and larger scale (e.g., Wikipedia categories or YAGO classes).
There, many categories have complex names that reflect human classification and organization, and as such, encode knowledge about class attributes, taxonomic, and other semantic relations~\citep{Nastase:2008:DWC, Pasca:2017:GTV}.
As a result, categories represent a unique and valuable resource, which has been exploited for various tasks, including entity retrieval~\citep{Ciglan:2012:SMA,Zhang:2017:ESA,Kaptein:2013:ECS}, query understanding~\citep{Balog:2011:QME}, and knowledge acquisition~\citep{Nastase:2008:DWC,Pasca:2017:GTV}. 

\begin{figure}[t]
   \centering
   \vspace*{0.5\baselineskip}
   \includegraphics[width=0.45\textwidth]{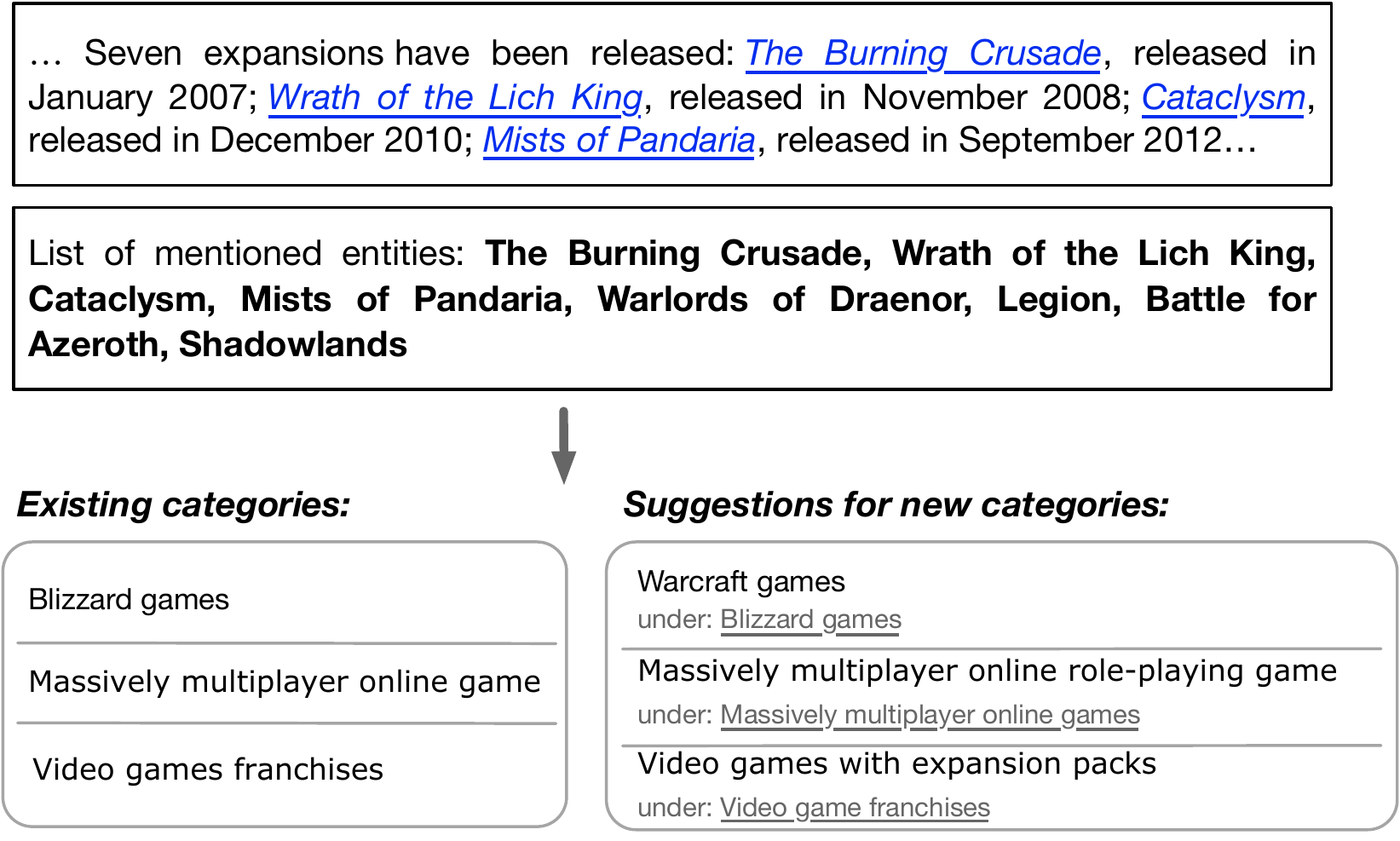} 
   \vspace*{-0.5\baselineskip}
   \caption{Given a set of entities along with context (surrounding text), we aim to create new categories, if no fine-grained enough category exists in the category system. 
   To ensure that a new category can meaningfully complement the current category system, we also find its place (i.e., parent category) in the hierarchy.  These suggestions are presented to a human editor for consideration.}
\label{fig:ill}
\vspace*{-\baselineskip}
\end{figure}

A large body of prior work focuses on assigning entities to categories, which is usually referred to as the task of  fine-grained entity typing~\citep{Obeidat:2019:DZF, Zhou:2018:ZOE, Choi:2018:UET}. 
Entity typing deals with a single entity as input (often with some surrounding text, as context), and it implicitly assumes that there exist suitable types/categories.
However, since new entities continuously  emerge~\citep{Hoffart:2014:DEE,Graus:2018:BCM,Zhang:NED:2020}, this assumption is unrealistic in practice, and brings about the need for creating new categories.\footnote{We note that it is a less severe issue for types, which are coarse-grained.}  The problem of expanding hierarchical category systems of entities with new categories has not received due attention to date.

An important consideration in this work is that category creation should not be driven by individual entities.  Looking at single entities may not be informative, as each could be an outlier.  On the other hand, when there is a set of entities for which no suitable, i.e., fine-grained enough, category exists, that would be a reason for expanding the category system.
Note that it is very natural to operate with sets of entities; lists and tables are ubiquitous on the Web.  Tapping into these would allow us to complement existing category systems.
Taking Fig.~\ref{fig:ill} as an illustration, we can recommend existing categories for this set of mentioned entities, but they are not fine-grained enough.
In this case, the creation of new categories would be desired.

The challenge we tackle in this paper is the following: given set of entities, generate new categories to meaningfully extend a given category system.
The main difference between ``plain'' entity sets and categories is that the latter are ``named'' sets which are organized hierarchically.
This gives rise to two specific novel sub-problems we are addressing in this work: (i) generating a label for the new category (comprising of the input set of entities), and (ii) finding its place in the category system (i.e., locating an appropriate parent category). 
%The former task resembles generating labels for a set of items, while the latter will help uncover the characteristics of the category system.

To study this problem, we take Wikipedia categories as a representative of a large-scale hierarchical category system that is widely utilized.
Currently, Wikipedia categories are created manually by editors (``Wikipedians''). 
With the increasing number of Wikipedia pages, more categories are created and added for organization. 
In 2012, there were about 660k categories, while at the time of writing there are over 1.1M categories. Nevertheless, the category system is still extremely sparse and noisy, as it contains duplications, errors, and oversights~\citep{Boldi:2016:CWC}.
Given the sheer number of categories, the maintenance and expansion of the category system are becoming increasingly difficult. It would be of great practical value to provide a method for automatically generating categories, triggered by updates made to a Wikipedia article.
The specific instantiation of our general problem in this application scenario is the following.
We assume a user is editing a Wikipedia article, where a list of entities has already been mentioned. 
Against this setting, our objective is to
see if new categories could be generated, based on the list of entities, which may be added to Wikipedia.  These suggestions are presented to a human editor for consideration.
More generally, we aim to generate new categories in an automatic manner for sets of entities.
While in this work we focus exclusively on Wikipedia categories, the methods we develop generalize to other category systems as well. 
% \todo{$\langle$UPDATE$\rangle$}

%
To control the quality of category labeling, we follow two general rules (in accordance with Wikipedia guidelines): (1) a label should be as specific as possible; and (2) similar categories should be avoided~\citep{Ma:2018:LFW}.
Inspired by this, we identify four main challenges related to the automatic generation of categories.
\begin{itemize}
	\item \textbf{(C1) Specificity:} A proper category should be sufficiently specific to capture the entity set intent.  For example, \emph{2018 Oscar Winners} is a more specific category than \emph{Actors}, although both might describe the contents of an entity set.
	\item \textbf{(C2) Hierarchy:} The more specific a category, the deeper down it is located in a hierarchy.  We explicitly focus on leaf categories and aim to place them in the hierarchy by finding their respective parent categories. 
	\item \textbf{(C3) Redundancy:} Redundant categories should be avoided.  While we are not addressing this problem explicitly in this paper, we postulate that by identifying parent or sibling categories, and presenting these to the human editor, would help address this issue. %
	\item \textbf{(C4) Importance:} An important category is expected to organize salient entities and encode knowledge, which is not already covered by sibling categories.  
\end{itemize}

\noindent
We propose a pipeline architecture consisting of several steps to overcome the above challenges. 
Specifically, given a set of entities as well as the surrounding text,
we aim to generate a ranked list of categories that capture the semantics of that set. 
Categories can be generated by summarizing the words contained in the input.
The ranked list may be further refined by incorporating signals from the context (e.g., surrounding text and page title). 
A particularly important subtask is to find the generated category's place in the hierarchy, by identifying its parent category. This information can then be further utilized as an additional ranking signal.
In summary, the contributions of this work are as follows.
\begin{itemize}
    \item We propose the task of generating categories for sets of entities, and develop an approach to generate categories that are specific, hierarchical, nonredundant, and important.
    \item We develop a test collection based on Wikipedia categories and lists, and perform an extensive evaluation of the proposed approach.
\end{itemize}
\section{Related Work}
\label{sec:bac}

The task of generating new categories from entity sets for extending the category system of a  knowledge base is related to the problems of entity typing, conceptual labeling, knowledge base population using semi-structured data, and ontology generation.

\begin{figure*}[!ht]
   \centering
   \vspace*{-0.5\baselineskip}
   \includegraphics[width=0.9\textwidth]{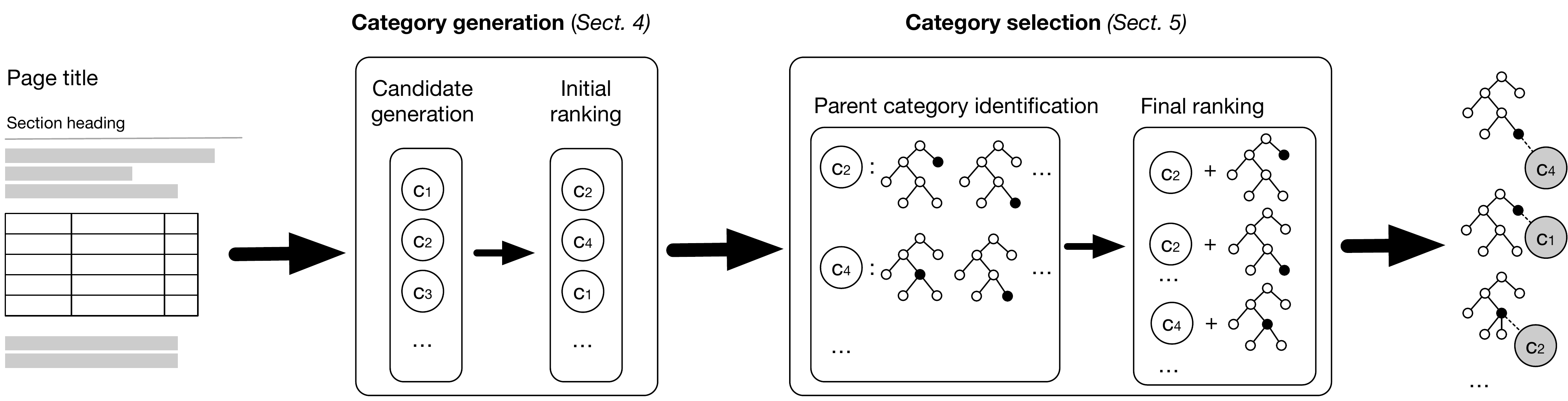} 
   \vspace*{-0.75\baselineskip}   
   \caption{Illustration of our category generation pipeline.}
   \vspace*{-0.5\baselineskip}
\label{fig:task}
\end{figure*}

\emph{Entity typing} refers to the task of assigning types to mentions of entities in context~\citep{Ren:2016:AFE, Obeidat:2019:DZF, Zhou:2018:ZOE, Choi:2018:UET}.
The roots of this problem may be traced back to named entity recognition (NER), which focuses on the detection of entity mentions in text and type-annotation of these mentions from a small set of coarse types (such as person, organization, location, and miscellaneous)~\citep{Nadeau:2007:SNE}.
Over the years, attention in NER has shifted from annotating against a small set of coarse types to using fine-grained types from a type taxonomy, which allows multiple and hierarchical types~\citep{Yosef:2012:HHT}. 
Unlike traditional NER work, recent entity typing approaches assume entity mentions to be provided as part of the input.
For example, AFET~\citep{Ren:2016:AFE} is a hierarchical partial-label embedding method for automatic fine-grained entity typing. It learns embeddings for mentions and types-paths, and iteratively refines the model.
\citet{Choi:2018:UET} introduce a new entity typing task to predict a set of free-form phrases, namely ultra-fine types, that describe the role the entity plays in a given sentence. 
Having free-form noun phrases as type descriptors, as opposed to existing fined-grained types from a taxonomy, is intended to improve downstream entity-focused tasks. 
Though both entity typing and our task deal with types and entities, we focus on creating new categories for entity sets, as opposed to dealing with a single entity. %which has not been considered in entity typing.

Another related problem is \emph{conceptual labeling}, which is the task of generating a small set of labels that best describe a set of words or phrases~\citep{Sun:2015:OCL}.  Labels can be concepts, entities, or types/categories from a knowledge base.  While conceptual labeling takes a set of items as input, its objective is language understanding with the help of existing concepts/types.  This is fundamentally different from our objective of expanding category systems of knowledge bases.

\emph{Knowledge base population} (KBP) typically involves two tasks, entity linking and slot filling~\citep{Ji:2011:KBP,Zhang:2020:WTE}. 
The former aims to link (often ambiguous) entity mentions in text to specific entries in a knowledge base, while the latter is concerned with completing the information available on a given entity. 
Sets of entities, as defined by tables and lists, have been considered in KBP.
For example, table-to-KB matching is considered as a fundamental step towards utilizing tables for KBP, and involves two specific sub-problems: entity linking for tables and table schema to predicate matching. 
\citet{Ritze:2015:MHT} propose an iterative method for matching tables to DBpedia. They develop a manually annotated dataset for matching between a Web table corpus extracted from Common Crawl~\citep{Lehmberg:2016:LPC} and DBpedia. In follow-up work, \citet{Ritze:2017:MWT} focus on a feature study for the same task. Specifically, they consider features extracted based on the tables themselves or from the knowledge base, and evaluate the utility of these for the matching task. %Above work focus on how to match table to KBs. 

Category creation is also related to the task of \emph{ontology generation}~\citep{Bedini:2007:AOG, Abeer:2015:OCT, Pivk:2005:AOG,Yu:2007:OEU,Sanderson:1999:DCH, Blei:2010:NCR,Stoica:2007:ACH}. 
It can be based on various sources, from unstructured free text to structured tabular data.
An example for the latter is the work by \citet{Pivk:2005:AOG}, who proposes automatic ontology generation using tabular structures. The steps include extracting and transforming a table into a regular matrix form, detecting table structure like orientation, schema and cells, identifying table type, rearranging table regions into a directed acyclic graph, and generating an ontology based on the graph.
\citet{McGuinness:2002:OCA} describes the scope of ontology specifications from simple ontologies to structured ones. The Wikipedia category structure corresponds the former because of its information hierarchy, but it is not a strict and logically grounded ontology due to the inconsistencies and loose relationships~\cite{Yu:2007:OEU}. 
%Additionally, it is a real application used by many users and more explicit comparing to ontologies.
There are two main distinguishing aspects of our approach. First, we expand an existing taxonomy, unlike most prior approaches that generate categories from top to bottom.  Second, we generate suggestions to assist a human editor instead of aiming for a fully automated category population.

The machine-readable semantic knowledge provided by taxonomies has proved to be beneficial in an array of natural language understanding problems~\citep{Biemann:2005:OLT}.  Wikipedia categories represent a unique and extremely valuable resource in this capacity, and have been utilized for a broad range of tasks~\citep{Ciglan:2012:SMA,Ma:2018:LFW,Kaptein:2013:ECS,Balog:2011:QME,Nastase:2008:DWC,Pasca:2017:GTV,Nastase:2013:TWL}.

\section{Problem Statement and Overview}
\label{sec:ps}

We provide a formal definition of the problem we are addressing.
\begin{mydef}[Category generation based on entity sets]
	Given a set of entities $E$ and its context $E_{\mathcal{C}}$, category generation is the task of generating a ranked list of category suggestions $\langle c_1,\dots,c_n \rangle$.  Each category suggestion $c$ is to be added as a leaf node in the category hierarchy under an existing parent category $c_p$.
\end{mydef}
We assume that the context of $E_{\mathcal{C}}$ includes the title of the page and of the context where the entities lie, as well as any existing categories that have already been assigned to that article (by a Wikipedian).

We address the category generation task using a pipeline approach, shown in Figure~\ref{fig:task}. The two main components of this pipeline are category generation (Sect.~\ref{sec:cg}) and category selection (Sect.~\ref{sec:gci}).
In \emph{category generation}, we first generate candidate categories that are relevant to the input entities and their context.
E.g., given a set of entities about ``European national teams not affiliated to FIFA,'' it generates candidates like \emph{National football teams in the isle of man}, \emph{Football teams in the isle of man}, \emph{Isle of man}, \emph{National association football teams} and \emph{European national and official selection-teams not affiliated to FIFA}.
Then, we perform an initial ranking of the candidates, using a set of inexpensive features based on structure and content information.

In the second pipeline component, \emph{category selection}, we first attempt to find the place of each candidate category in the category hierarchy by locating possible parent categories.
For instance, for a candidate category \emph{European national and official selection-teams not affiliated to FIFA}, possible parent include \emph{National and official selection-teams not affiliated to FIFA}.
Then, we perform a final ranking of the candidates, using their predicted place in the hierarchy as an additional ranking signal.  This allows us to more accurately estimate category importance as well as to avoid creating redundant categories. In the above example, the candidate \emph{National association football teams} would be excluded.
Finally, the top-ranked suggestions would then be presented to a human editor, who can decide which of these categories, if any, should be created.

\section{Category Generation}
\label{sec:cg}

The first component of our pipeline is responsible for the generation of candidate categories.  In particular, in this part, we address the challenge of \emph{specificity} (\textbf{C1}).
%To overcome specificity, we address the task of category generation, 
Our approach consists of a candidate generation step based on pointer-generator networks (Sect.~\ref{sec:cg:cg}), followed by an initial ranking of the candidates, using a set of inexpensive features (Sect.~\ref{sec:cg:ir}). 

\subsection{Candidate Generation}
\label{sec:cg:cg}

Abstractive summarization models, which are used for summarizing documents into a few sentences, have proved effective to generate titles for the semi-structured data~\citep{Braden:2019:GTW}. In this work, we consider three neural text summarization models to learn to generate categories.
We concatenate all the entities and contextual data into key-value pairs and use them as the representation of input data to be fed into the summarization models. % to generate new categories.

\begin{itemize}
	\item \emph{Pointer-generator network}~\citep{See:2017:GTP} is a hybrid neural network that combines pointing and generating mechanisms.\footnote{The network allows to copy words via pointing as well as generate words from a fixed vocabulary.}
It uses a bi-LSTM encoder and LSTM decoder with attention. 
The generation probability $P_{g} \in [0,1]$ for each step is calculated from the context vector $c_t$, the decode state $h_t$, and the decode input $x_t$:
\begin{equation}
	p_{g} = \sigma (W_c \cdot c_t + W_h \cdot h_t + W_x \cdot x_t + b) ~,
\end{equation}
where $W_c$, $W_h$ and scalar $b$ are learnable parameters, and $\sigma$ is the sigmoid function.
$P_g$ is used as soft switch to choose to sample from the vocabulary distribution $P_v$ or to copy a word from the input with the attention distribution $P_a$.
The loss for each token $w$ is:
\begin{equation}
	P = P_g \cdot P_v(w) + (1-p_g) \cdot Pa (w) ~.	
\end{equation}
The loss function is the average negative log likelihood of the generated sequence.
In addition, \citet{See:2017:GTP} propose a coverage mechanism to overcome token duplications; here, we take the pointer-generator itself as the base model.

\item \emph{NATS}~\cite{Shi:2018:NAT} is an abstractive text summarization approach that extends the pointer-generator network.  One issue summarization models suffer from are repetitions (both word-level and sentence-level). To overcome it, Intra-decoder~\citep{Paulus:2017:DRM} allows a decoder to keep track of previously decoded tokens apart from the source data, not repeatedly producing the same information.
Additionally, sharing the weights with the decoder is a common solution that can boost performance. 
In summary, we take the pointer-generator network as the base model and equip it with the coverage mechanism and unknown words replacement, as in~\cite{Shi:2018:NAT}. % added from the literature by NATS.

\item \emph{FAST}~\citep{Chen:2018:FAS} is a neural generative model that first selects salient sentences and then rewrites them abstractively. Making use of salient information from the extraction process is another way to improve the summarization~\citep{Shi:2018:NAT}. 
The FAST model consists of an extractor agent and an abstractor network. 
The extractor aims to learn to extract the salient data from the source by training a pointer network, while the abstractor rewrites the extracted sentences to get the final summarization.

\end{itemize}

\begin{table*}[!t]
\centering
\caption{Features for category classification.  Initial ranking (in Sect.~\ref{sec:cg:ir}) uses the first two blocks of features, while final ranking (in Sect.~\ref{sec:gci:fr}) uses all features.}
%\vspace*{-0.5\baselineskip}
\label{tbl:features}
\begin{tabular}{lllc}
	\toprule
	& \textbf{Feature} & \textbf{Explanation} & \textbf{\#Features} \\
	\midrule
	\multicolumn{4}{l}{\emph{I. Structure-based features}} \\
	\midrule
	& $|c|$ & Length of the category (number of terms) & 1\\
	& $\mathit{IsPrepos}(c)$ & Binary indicator whether $c$ contains prepositions & 1\\
	& $\mathit{IsStopwords}(c)$ & Binary indicator whether $c$ contains stopwords & 1\\
	& $\mathit{IsEntity}(c)$ & If $c$ is an entity & 1\\
	& $|\mathcal{C}_e|$ & Number of categories if $c$ is an entity &1\\
	& $\mathit{CatNum}_{\mathit{aggr}}(E_c)$ & Aggregation of the number of categories for entities contained in $c$ & 3\\	
	& $\mathit{TermFreq}_{\mathit{aggr}}(\mathcal{T}_c)$ & Aggregation of term frequencies for terms in $c$ & 3\\
	\midrule
	\multicolumn{4}{l}{\emph{II. Content-based features}} \\
	\midrule
	& $\textit{CatNameSim}(c,\mathcal{C})$ & Top-$k$ BM25 scores using category labels & $k$ \\
	& $\textit{EntityNameSim}(c,\mathcal{E})$ & Top-$k$ BM25 scores using entity labels & $k$\\
	& $\textit{EntitySim}(\mathcal{E}_c,\mathcal{E}_{c'})$ & Top-$k$ BM25 scores using member entities & $k$ \\ 
	& $\textit{EntityOverlap}(\mathcal{E}_c,\mathcal{E}_{c'})$ & Top-$k$ member fractions against input category ($|\mathcal{E}_c \cap \mathcal{E}_{c'}| / |\mathcal{E}_{c}|$) & $k$ \\
	& $\textit{EntityOverlap2}(\mathcal{E}_c,\mathcal{E}_{c'})$ & Top-$k$ member fractions against candidate category ($|\mathcal{E}_c \cap \mathcal{E}_{c'}| / |\mathcal{E}_{c'}|$) & $k$\\
	& $\textit{ParentCatSim}(\mathcal{C}_c,\mathcal{C}_{c'})$ & Top-$k$ BM25 scores using categories of $c$ & $k$\\ 
	& $\textit{ParentCatOverlap}(\mathcal{C}_c,\mathcal{C}_{c'})$ & Top-$k$ category fractions against input category ($|\mathcal{C}_c \cap \mathcal{C}_{c'}| / |\mathcal{C}_{c}|$) & $k$\\
	& $\textit{ParentCatOverlap2}(\mathcal{C}_c,\mathcal{C}_{c'})$ & Top-$k$ category fractions against candidate category ($|\mathcal{C}_c \cap \mathcal{C}_{c'}| / |\mathcal{C}_{c'}|$) & $k$ \\
	\midrule
	\multicolumn{4}{l}{\emph{III. Category importance features}} \\
	\midrule
    & $\mathit{Importance}_{\mathit{aggr}}(c)$ & Aggregation of the number of categories containing each segment of the category name & 3\\
    & $\mathit{Inlinks}_{\mathit{aggr}}(c)$ & Aggregation of the numbers of inlinks of member entities & 3\\
    & $\mathit{Outlinks}_{\mathit{aggr}}(c)$ & Aggregation of the numbers outlinks of member entities & 3\\
%    & $\mathit{Changes}(c_p)$ & Num/rate of member entities or parent categories deleted & 3\\
%    & & Num/rate of categories deleted && FR\\
    & $\mathit{GraphSize}(c, c_p)$ & Total number of member entities, siblings and parent categories & 1\\
    & $\mathit{GraphStat}(c, c_p)$ & Aggregation of the number categories of member entities in the parent category ($c_p$) & 3 \\
%    && \sz{add more features here} \\
	\bottomrule
%	\multicolumn{3}{l}{$\dag$ Three features, one for each aggregator (max, sum, avg)} \\
%	\multicolumn{3}{l}{$\ddag$ $k$ features} \\
\end{tabular}
\end{table*}

%\vspace*{-0.25\baselineskip}
\subsection{Initial Ranking} 
\label{sec:cg:ir}

We perform an initial ranking of candidate categories using structure-based and content-based features.  These features are fed into a Random Forest regressor in order to obtain an initial ranking of the candidates. Only the top-$k$ ranked candidates are kept for downstream processing.

\subsubsection{Structure-based Features}
The first set of features is based on structure (or patterns) in the category names; these are listed in the top block of Table~\ref{tbl:features}. 
%It reveals the characteristics of category structures. 
Most features are simple characteristics.
The first one is the length of the category name ($|c|$).
The next feature ($\mathit{IsPrepos}(c)$) is motivated by the observation that prepositions are a strong indicator of semantic relations~\citep{Lauer:1996:DSL}. E.g., \emph{Films directed by Joss Whedon} indicates the explicit relation between \emph{Films} such as \emph{The Avengers} and \emph{Joss Whedon}. We thus consider the presence of a preposition as a feature. 
Further, we decompose category names based on entities mentioned in them and write $E_c$ to denote the set of entities in category $c$.  We find that over 91\% of Wikipedia categories contain entities.
%For instance, \emph{2018 in Japan} contains the entities of \emph{2018} and \emph{Japan} or \emph{Intracoastal Waterway} is both an entity and a category. 
Then, two features are constructed based on $E_c$, namely,
the number of entities $|E_c|$, and the aggregation on the number of categories of the elements in $E_c$ according to:
%
%%
%or the number of entities contained in the candidate name, or the number of existing categories that overlap somehow with the candidate name. }
%
%For example, the category length. Wikipedia categories are entity-oriented, and over 91\% of them embed entities inside the labels. 
%In many cases, categories are entities, e.g.
%``Intracoastal\_Waterway'' is both an entity and Wikipedia category.
%The category size reflects the impact of an entity. So if $c$ is an entity, we take the number of its categories $|C\_e|$ as one feature. If $c$ embeds entities, e.g., ``2018\_in\_Japan'' embeds the entities of ``2018'' and ``Japan'', we aggregate the number of categories of the embedded entities ($\mathcal{E}_c$) as features:
%
\begin{equation}
        \mathit{CatNum}_{\mathit{aggr}}(E_c) = \mathit{aggr}\big( \{|C_e|: e \in E_c\} \big )~,
\end{equation}
where $C_e$ indicates the set of categories entity $e$ is a member of and $\mathit{aggr}$ is an aggregator function. Specifically, we use $\mathit{max}$, $\mathit{sum}$, and
$\mathit{avg}$ as aggregators.  Similarly, we also take the aggregated category term frequencies as signals.

\subsubsection{Content-based Features}

The second set of features, listed in the middle block of Table~\ref{tbl:features}, are based on the content of categories, measured in terms of labels, entities contained, and parent categories.
%The content listed in a Wikipedia category page includes its member entities (Wikipedia pages) organized under it, sub-categories and parent categories. We propose a number of search-based features to investigate the content-wise consistency. 
One group of features is based on the similarity of category labels. Intuitively, a new category should be consistent with the naming of existing categories.
Using the candidate category name as a keyword query, we rank all existing categories $\mathcal{C}$ in Wikipedia using BM25, and take the retrieval scores of the top-$k$ highest ranked categories as features ($\textit{CatNameSim}(c,\mathcal{C})$). 
Similarly, we also rank all entities $\mathcal{E}$ in Wikipedia using the candidate category name as the query and use the top-$k$ scores as features ($\textit{EntityNameSim}(c,\mathcal{E})$). 

We can also measure the similarity between the candidate category $c$ and existing categories $c' \in \mathcal{C}$ in terms of the entities they contain.  For this, we operate on an index of categories where tokens are member entities.  To estimate the member entities of the candidate category $\mathcal{E}_c$, we take all entities that are present in the input set.  Then, we create a search query by enumerating all entity tokens in $\mathcal{E}_c$ to rank existing categories $c' \in \mathcal{C}$.  The top-$k$ retrieval scores are used as features ($\textit{EntitySim}(\mathcal{E}_c,\mathcal{E}_{c'})$).  Additionally, we compute the entity overlap between $\mathcal{E}_c$ and each of the top-$k$ existing categories $\mathcal{E}_{c'}$ in two different ways ($\frac{|\mathcal{E}_c \cap \mathcal{E}_{c'}|}{|\mathcal{E}_{c}|}$ and $\frac{|\mathcal{E}_c \cap \mathcal{E}_{c'}|}{|E_{c'}|}$).

%Suppose we have known the entities members and categories of c, we compare them against the existing categories to avoid duplications. For instance, \emph{War of the Worlds comics} and \emph{Comics based on The War of the Worlds} share most of the entity members and parent categories, and only the latter is kept for the current Wikipedia. 
%We calculate three types of scores as features. The first one is $\textrm{score}_1(c,c')$, which is a BM25 score. $\textrm{score}_2(c,c')=
%and $\textrm{score}_3(c,c')=\frac{|E_c \cap E_{c'}|}{|E_{c'}|}$
%, where $E$ denotes the member entities, $\textrm{score}_2$ indicates if there is already a category $c'$ organizing the same group of entities, and $\textrm{score}_3$ tells the likelihood that c is split or merged with $c'$. 

Finally, we use parent categories ($\mathcal{C}_c$ and $\mathcal{C}_{c'}$), analogously to member entities, by creating an index of categories where tokens are their parent categories (i.e., treating categories as atomic units, without splitting up their labels) and computing the same features.

To rank candidates, all structure- and content-based features are fed into a Random Forest regressor. % (see Sect.~\ref{sec:tc:cg} for details).

%
%\begin{equation}
%P(c) = \lambda P_{sc} + (1-\lambda) \cdot P_{Jaccard}(c, \mathcal{C}_{\mathcal{T}}),	
%\end{equation}
%where $P_{sc}$ is the classification score using structure and content features, and $P_{Jaccard}$ is the Jaccard similarity between $c$ and the contextual text.

\section{Category Selection}
\label{sec:gci}

Next, we select categories that are deemed appropriate, by ranking the candidates generated in the previous section, then pruning the ranked list.
Specifically, we start by identifying the potential parents of each candidate category (Sect.~\ref{sec:gci:pci}).  Our final ranking then considers the place of the candidate category in the hierarchy (Sect.~\ref{sec:gci:fr}), thereby addressing the challenges of \emph{hierarchy} (\textbf{C2}), \emph{redundancy} (\textbf{C3}), and \emph{importance} (\textbf{C4}).

% 5.1 hierarchy and redundancy
% 5.2 importance

\subsection{Parent Category Identification}
\label{sec:gci:pci}

An important element of our approach is to find the place of the candidate category in the hierarchy.  This is cast as a ranking problem: given a candidate category, return a ranked list of possible parent categories. 

It is intuitive to think that a category would be named similar to its parent categories. %in terms of topic and content.  
Indeed, we find that nearly 90\% of the category-parent ($\langle c,c_p \rangle$) pairs in Wikipedia share at least one term, %\sz{where $\langle c^{WP},c_p^{WP} \rangle$ denotes all Wikipedia category and parent category pairs}. 
e.g., $\langle$\emph{Companies established in 1974},  \emph{Clothing companies established in 1974}$\rangle$.  However, there are also numerous examples of categories that share terms with many other non-related categories, e.g., \emph{Dragons} (mythological monsters) and \emph{Dragon age} (fantasy role-playing game).  In other cases, the category and its parent only share the topic, but not any of the terms, e.g., $\langle$\emph{Middle-earth Valar}, \emph{Fictional deities}$\rangle$.
Additionally, the Wikipedia category system suffers from the violation of the transitivity principle, i.e., a category may contain irrelevant subcategories~\citep{Kirillovich:2018:OAO}, e.g., $\langle$\emph{Flight}, \emph{Motion (physics)}$\rangle$. 
Simple term-based matching techniques are thus unlikely to be sufficient, making it an extremely challenging task.

To overcome the above issues, we construct a topic graph based on all (existing) $\langle c,c_p \rangle$ pairs in Wikipedia.  Inspired by~\citet{Lawrie:2001:FTW}, we aim to find topic words for multi-document summarization. 
The model is composed of all conditional probabilities $P(t_a|t_b)$, where $t_a$ is a topic term in $c_p$ and $t_b$ is a category term in $c$.  For simplicity, we will be referring to terms, noting that these can also be multi-word phrases.
We segment the categories by proposition.
For example, the parent category of \emph{1903 establishments in Colombia} is \emph{1903 establishments in South America}, and we segment them to $\{$\emph{1903 establishments, South America}$\}$, and $\{$\emph{1903 establishments, Colombia}$\}$ respectively, and take \emph{South America} as $a$ and \emph{Colombia} as $t_b$.  We segment categories by tokens for those that do not have prepositions.
Then, we set
\begin{equation}
	P(t_a|t_b) = \frac{n(t_a,t_b)}{n(t_b)} ~,
	\label{eq:pab}
\end{equation}
where $n(t_a,t_b)$ is the number of $\langle c,c_p \rangle$ pairs where $c_p$ contains $t_a$ and $c$ has $t_b$, and $n(t_b)$ is the number of categories containing $t_b$. 
We form a graph by considering each parent topic term and category term as vertices, and assign $P(t_a|t_b)$ as the weight of the edge between them.

\begin{figure}[t]
%   \vspace*{-0.75\baselineskip}
   \centering
   \includegraphics[width=0.45\textwidth]{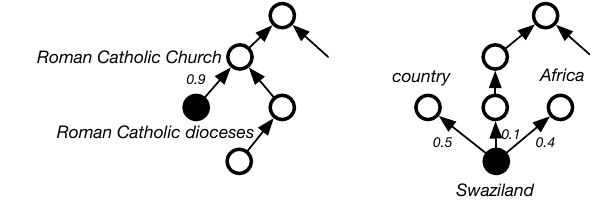} 
%   \vspace*{-0.75\baselineskip}
   \caption{Excerpt from the topic graph, which is utilized both for query expansion and for ranking parent categories. When used for query expansion, the selected topic words for a candidate category \emph{Swaziland} would include \emph{country} and \emph{Africa}. When used for ranking, $P(t_a|t_b)$ between \emph{Roman Catholic Church} and \emph{Roman Catholic dioceses} is 0.9.}
\label{fig:topic}
\end{figure}

This topic graph, illustrated in Fig.~\ref{fig:topic}, is then leveraged in two orthogonal ways: for query expansion and for ranking.

\subsubsection{Query Expansion}
\label{sec:cs:qe}

The name of the category $c$ can be used as a keyword query to find related categories.  Additionally, we propose a query expansion mechanism to generate an expanded query. We leverage the topic graph to add the top topic parent terms. Instead of using a fixed rank-based cutoff, we employ dynamic thresholding.  Specifically, we set $\frac{1}{\alpha} \max_{t_a} P(t_a|t_b)$, where $t_b \in c$, as the threshold.  Notice that this ensures that at least one topic term is always selected. 
Terms with weights exceeding the threshold are chosen and put in the topic set $\tilde{T}_t$.  
Terms in the set will form a new query by appending the topic words to the original query, i.e., 
%(2) refine the category by replace the category with the topic term, e.g., \emph{Scandinavian} countries will be refined as \emph{Northern European} countries.
%
\begin{equation}
	\tilde{c} = c \cup \tilde{T}_c ~,	
	\label{eq:queryexp}
\end{equation}
where $\tilde{T}_c = \cup_{t \in c} \tilde{T}_t$ is the entire set of the topic parent terms of $c$.
%See Fig.~\ref{fig:topic} for an example.  
Then, %either $c$ or 
$c$ or $\tilde{c}$ is used as a keyword query to retrieve parent categories.

\subsubsection{Ranking using Topic Graph}
Given a keyword query $q$, which may be the category name $c$ or the expanded query $\tilde{c}$ (cf. Eq.~\eqref{eq:queryexp}), 
we propose the following \emph{hierarchy-based} retrieval model based on the topic graph: 
\begin{equation}
	\mathit{\phi}_{H}(q, c_p) = \frac{\sum_{t_a \in c_p} \sum_{t_b \in q} P(t_a|t_b)}{|c_p|\cdot|q|} ~,
\end{equation}
%
%where $c_p$ denotes the candidate parent category.
%
%Given a query, we consider three retrieval methods:
%BM25, $P(c_p|c)$ and \sz{combining} BM25 with $P(c_p|c)$. $P(c_p|c)$ is the averaged topic weights based on Eq.~\ref{eq:pab}:
%\begin{equation}
%	P(c_p|c) = \frac{\sum_{a \in c_p} \sum_{b \in c} P(a|b)}{|c_p|\cdot|c|}.
%\end{equation}
%Here $c_p$ denotes a \sz{parent} category.
%We name the last ranking method as \sz{hierarchy}-based retrieval model, and express it as follows.
where $\phi$ stands for the retrieval score, and $|c_p|$ and $|q|$ are measured in the number of segmented terms.
%\todo{say that $|c|$ measured in the number of segmented terms}
This hierarchy-based model can then be combined with a content-based retrieval model (here: BM25) in a simple multiplicative manner:
\begin{equation}
	\mathit{\phi}_{\mathit{combined}}(q,c_p) = \mathit{\phi}_{BM25}(q,c_p) \cdot \mathit{\phi}_{H}(q,c_p) ~.
\end{equation}

\subsection{Final Ranking}
\label{sec:gci:fr}

The final ranking step considers the place of each category in the hierarchy.  Given a candidate category $c$ and the set of its top-$k$ possible parent categories $C_p$, we score each $\langle c, c_p \rangle$ pair ($c_p \in C_p$), denoted as $\psi$.  Then, the final score for each candidate category is computed using:
\begin{equation}
	\mathit{score}(c) = \max \{\mathit{\psi}(c,c_p)| c_p \in C_p\} ~.
\end{equation}
The above score determines the order in which suggestions are presented to Wikipedia editors. Additionally, we filter out low quality suggestions, whose score is below a pre-defined threshold (i.e., $\mathit{score}(c) < \gamma$).

To estimate $\mathit{\psi}(c,c_p)$, 
%\kb{[This is somewhat confusing as we use $\mathit{score}(c,c_p)$ in the previous subsection, but this refers to a different score]}
we use the same feature-based supervised learning approach as for the initial ranking (cf. Sect.~\ref{sec:cg:ir}), but introduce a third group of features. These \emph{category importance features} are listed in the bottom block of Table~\ref{tbl:features}. 
%
%Categories are created for organizing Wikipedia pages, and the category importance estimates its impact. E.g., \emph{Brightline templates} contains only one Wikipedia entity member and it is being considered for deletion due to the decreasing importance. We propose a few features estimating the category importance. 
We assume that the importance of a category depends on its content.
%, and especially the ``weakest'' content. For example, \emph{Science and technology in Massachusetts} is important as both \emph{Science and technology} and \emph{Massachusetts} are \sz{important}. However, \emph{Trees in Massachusetts} is less important because \emph{Trees} of an institute is a less interesting property. 
To capture content importance, we segment the candidate category into the set $\mathcal{S}$ using the following rules: prepositions are isolated, longest prefix and suffix are kept if they can be found in other categories.  E.g., \emph{Geography of North Yorkshire} is segmented as $\{$\emph{Geography}, \emph{of}, \emph{North Yorkshire}$\}$ %instead of $\{$\emph{Geography}, \emph{of}, \emph{North}, \emph{Yorkshire}$\}$ 
(as \emph{North Yorkshire} is the longest prefix found in other categories before a preposition).  %It is under the pattern of \emph{X of Y}. 
We estimate a category's importance by aggregating those of its segments:  %, e.g., \emph{X} and \emph{Y}. we use the number of categories containing \emph{X} or \emph{Y} to represent the importance. It is expressed as follows.}
\begin{equation}
    \mathit{Importance}_{\mathit{aggr}}(c) = \mathit{aggr}_{s \in \mathcal{s}} \{n(s)\} ~,
\end{equation}
where $n(s)$ is the number of (existing) categories that contain the segment $s$ in their name and $\mathit{aggr}$ is an aggregator function ($\mathit{max}$, $\mathit{sum}$, or $\mathit{avg}$).
Alternatively, category importance can also be characterized by hierarchical properties.
Since we know the corresponding parent category $c_p$, we can consider the number of member entities $\mathcal{E}_c$ of the candidate category (as before, approximating it by taking all entities that are present in the input set), number of siblings, as well as the size of other categories that the member entities $\mathcal{E}_c$ belong to.
%For example, the number of entities, parent categories, and aggregation of the number of siblings and entities' category size.
Apart from the number of member entities, their importance also matters. We estimate the member entities' importance by their number of Wikipedia inlinks and outlinks.
%We also check the number or rate of member entities and categories deleted from a previous snapshot of Wikipedia.
%\todo{$\Leftarrow$ Is this about the parent category? Because this info you don't have for the candidate category.}

\section{Experimental setup}
\label{sec:tc}
\if 0
\todo{
\begin{itemize}
    \item Highlight this as a novel contribution (new task, no test collection yet)
    \item Argue why this is realistic and not to narrow
    \item Also be explicit about potential limitations
\end{itemize}
}
\fi

We describe the creation of purpose-built test collections and detail our experimental setup.

%    \item Tables table multi-column data while lists is single-column data. In order to incorporate both tables and lists, we only consider single column for tables.
%    \item Wikipedia Lists 10.2017. Pagetitle starts with ``list of'' : 244625; Pagetitle+section: 264659; Pagetitle+section+caption: 293275.
%

\subsection{Test Collections}
\label{sec:data:tc}

%\paragraph{Dataset of Category Generation}
%\kb{This part is a core contribution; it need more details and possibly an illustration.}
%\sz{There are no publicly available test collections for this task. We create this dataset with a strategy sampling method, which deems as one of the contributions of this work.}

Since no test collection exists for our task, we need to develop evaluation resources for the end-to-end task as well as for specific components of our pipeline.  
%These resources represent an important contribution and will be made publicly available upon acceptance.

\subsubsection{End-to-end Evaluation}
\label{sec:data:tc:end-to-end}

Recall that our input is a set of entities in a Wikipedia page, and the output is a ranked list of category suggestions, presented to a Wikipedia editor (cf. Fig.~\ref{fig:ill}).
Rather than in random paragraphs, entity sets are more likely to appear in semi-structured formats such as tables and lists~\citep{Zhang:2019:ADC}.
We sample tables/lists from a Wikipedia dump as inputs and try to heuristically recover the categories that could be created based on the set of entities contained in them.  
We do so by leveraging the categories that are associated with the corresponding Wikipedia page.
% \sz{This way, to collect entities sets containing similar categories, is more realistic. 
% It additionally assists in finding the corresponding descriptive context of entity sets and therefore does not narrow to tabular data. 
% We do not hide the potential limitations, such as shadowed topic coverage.
% }

Specifically, our test collection consists of over 10k input tables/lists that are sampled from Wikipedia.
% the WikiTables corpus (based on a Wikipedia dump taken in October 2017). 
We limit ourselves to tables/lists that contain a reasonable amount of information, that is, have at least five entities, which is the entity set $E$. 
%(we focus on tables that are more likely to generate popular categories~\citep{Zhang:2017:ESA} 
%\todo{$\Leftarrow$ not sure about this argument; we are not limiting ourselves to relational tables, so why do relations matter?}).
For each input entity set $E$, we obtain the ground truth categories based on the categories that are assigned to the embedding Wikipedia page.  For each of the page's categories $c$, we check if over half of the entities in $E$ are members of that category.  If yes, then $c$ is added to the ground truth, i.e., is a good suggestion.
Then, category $c$, along with all its subcategories, is removed from the page as well as from the Wikipedia category system.\footnote{We manually exclude a handful of general categories, like \emph{Living people}.}  Our aim will thus be to ``rediscover'' $c$ based on the input entities (The remaining categories of the page will be utilized as part of the input.)
%\todo{$\Leftarrow$ it was removed from both the page and WP categories, right? :)}
We avoid ``trivial'' categories, that is, when the name of the category is the same as the title of the corresponding Wikipedia page.
Further, we make sure that no pair of entity sets have identical ground truth (so as to avoid ``leakage'' between training and test data).
Our test collection comprises of 10,542 tables/lists, originating from 10,149 Wikipedia pages (a page might contain multiple tables/lists).  On average, there are 1.55 correct category suggestions for entity set in the ground truth.
Finally, the test collection is split 80/10/10 into train/validation/test splits.

\subsubsection{Category Ranking}
\label{sec:data:tc:cr}

The category ranking component is used in both the initial and final ranking steps of our pipeline (cf. Sects.~\ref{sec:cg:ir} and~\ref{sec:gci:fr}).  To train a machine-learned model, we require a set of positive and negative category examples.
%We rank the generated candidates in the initial ranking and final ranking steps, respectively, which differ at the input.
While the former is straightforward, the selection of negative (``bad'') categories is challenging as those categories are non-existent.  This is the very fact we exploit: categories that existed for a while but got removed from Wikipedia are bad ones, while those that still exist are likely to be good ones.
%\sz{To construct the dataset for training the models,
%we sample 50k positive categories and 50k negative categories.
%We assume the existence is a signal to indicate the importance.
%Categories, which were created a few years ago and still exist nowadays, are good ones. 
%Categories, which only existed for a while, are bad ones.
Thus, we take snapshots of the Wikipedia category system at three different points in time (2012, 2016, and 2019) and check the existence of a given category across them. 
%\todo{[BEGIN simplify] In this part, I don't see the point of having 3 snapshots. If it matters, then just say that we consider three snapshots from years YYYY, YYY, and YYY and explain accordingly (without mentioning specific DBpedia versions).}
%We check their existences by referring to different versions of category systems.
If a category is present in all three snapshots (2012, 2016, and 2019), it is believed to be a sound category (positive example) given its long-lasting existence. 
In contrast, if a category exists in the 2012 snapshot, but does not appear in both 2016 and 2019, then it is deemed to be a poor category (negative example).
%\todo{[END simplify]}
%
We sample a total of 50k positive and 50k negative categories. 
From this set, we remove categories that had their labels updated or replaced (by Wikipedians) because of naming conventions.  E.g., \emph{Speakers of the National Assembly of Mauritius} was replaced with \emph{Speakers of the National Assembly (Mauritius)}.  We detect such changes using a set of simple rule-based methods (based on edit distance and member entity overlap). 
We further subsample 5k negative and 5k positive categories from the remaining categories.
Then, the resulting 10k categories are used as training data for the category ranker. % train the initial ranking ranker.
%Final ranking additionally considers a parent category as the input.}
Note that some features also consider the parent category (cf. Table~\ref{tbl:features}).  Since some categories have multiple parents, we select a single parent category that is the largest one (i.e., has the most member entities).

\subsubsection{Parent Category Identification}
\label{sec:data:tc:pgi}

Additionally, we create a separate test collection for evaluating the parent category identification component. 
We randomly sample 5k leaf categories as input, and take their corresponding parent categories (2.4 on average) as the ground truth.

\subsection{Experimental Setup}
\label{sec:eva:es}

\subsubsection{Evaluation Measures}

Both the end-to-end task and the various components are evaluated using standard rank-based measures (NDCG, MAP, MRR, and Precision) at cut-off $k$.  
%\sz{We consider different metrics for the components.
%For candidate generation, to estimate whether we can reproduce the existing categories, we} consider Normalized Discounted Cumulative Gain (NDCG) as the main metric.
%Additionally, we evaluate the generated category names in terms of Recall-Oriented Understudy for Gisting Evaluation (ROUGE), which is widely used in automatic summarization. 
%
%For parent category identification, we use Mean Average Precision (MAP) as our main evaluation metric. We also report on Mean Reciprocal Rank (MRR). 
%We keep the same metrics as the initial ranking for the final ranking step and additionally consider P@5 and MRR. 
%
We measure statistical significance using a two-tailed paired t-test, with Bonferroni correction.  We use $\dag$/$\ddag$ to denote significance at the 0.05 and 0.01 levels, respectively.

\subsubsection{Category Generation}
\label{sec:tc:cg}

We employ three candidate generation methods.
The pointer-generation network is based on~\citep{See:2017:GTP}, following the settings used in~\citep{Braden:2019:GTW} (which is used to generate titles for Web tables). 
For NATS~\citep{Shi:2018:NAT}, we use their publicly released toolkit.\footnote{https://github.com/tshi04/NATS}
We leverage the features of the coverage mechanism and unknown words replacement apart from the pointer-generator network.
We train the model for 35 epochs, and set the learning rate to 0.0001.
The FAST model is based on ~\citep{Chen:2018:FAS}. 
Word embeddings with 128 dimensions are generated using gensim\footnote{https://github.com/RaRe-Technologies/gensim} for the training data. We use the Adam optimizer and the same learning rate as for NATS for training the full network.
We perform beam search for all three approaches to choose the suitable models.  
The categories generated by the three models are combined in a candidate pool.

In the initial ranking step, we set $k=5$ (for content features that consider the top-$k$ similarity scores).
We train a regression model based on the 10k samples in our category ranking dataset (cf. Sect.~\ref{sec:data:tc:cr}) and apply it to rank the categories in the candidate pool. 
%\todo{$\Leftarrow$ is it correct this way?}
Specifically, we employ the Random Forest algorithm, with the number of trees set to 1000 and the maximum number of features in each tree set to 10.
Note that the initial ranker uses only the structure-based and content-based features in Table~\ref{tbl:features}.  %And the classification scores are used to initially rank the candidates in the candidate pool.

\subsubsection{Category Selection}

The second component of our pipeline comprises of two steps: parent category identification and final ranking.
For parent category identification, we build the topic graph using all the $\langle c,c_p \rangle$ pairs in Wikipedia, excepting the 5k test categories and their parents which we sampled for evaluation (cf. Sect.~\ref{sec:data:tc:cr}).
Next, we fetch the top 10 parent categories using query expansion.  We set $\alpha$ to 2 based on a set of preliminary experiments.
%We only take the entities from the tabular data that are sorted by $c$ as the members. \todo{$\Leftarrow$ why do we need this sentence? Not sure why we should talk about member entities for parent category identification.}

%\todo{Simplify this part. First, we discuss the final ranker. Then, explain that additionally, we employ BERT as a baseline (but no need for a separate subsubsection, just a new paragraph).}
For final ranking, we use all features in Table~\ref{tbl:features} and train a ranker with the same settings as for the initial ranking step.  We filter out low quality suggestions, using a score threshold of $\gamma = 0.1$.

%\todo{Not sure why this is here; integrate with the BERT part below.}

%Categories that are labeled as negative are excluded from the pool with a threshold of 0.1 and the rest are re-ranked using the classification scores.

%\subsubsection{BERT Baseline}
%\label{sec:tc:bert}

%\todo{Explain that we use BERT as a baseline for the the final ranking step. Also say that the potential data leakage issue (the category we want to generate may exist in the training data), thus BERT may have an unfair advantage.}
Additionally, we consider BERT~\citep{Devlin:2018:BPD} as a baseline for the final ranking step. 
%\todo{It's for final ranking, not end-to-end, right? If yes, then no need to discuss the additional components like parent category identification.}
%BERT is trained for a masked language model and next-sentence prediction tasks.  
BERT is a highly effective language representation model, which has been designed to be pre-trained from unlabeled text.  We take a pre-trained BERT model (``bert\_uncased\_L-12\_H-768\_A-12,'' which has been trained on the English Wikipedia and the BookCorpus) and fine-tune it for a \emph{sentence pair classification} task.
%In a Wikipedia category page, \sz{the category and its parent categories are placed in the same page}. 
%We consider BERT as the baseline by fine-tuning the task in this flavor: \emph{sentence pair classification}.
%We compose a sentence pair following a Wikipedia category page, i.e., given a $c$ as the first sentence, the second sentence is its parent category. 
Specifically, we compose sentence pairs by taking a Wikipedia category $c$ as the first sentence and its parent category $c_p$ as the second sentence.  We utilize the method in Sect.~\ref{sec:gci:pci} to find the parent categories, and use the sentence pair classification score for ranking. 
%\todo{Update notation, it's not $P()$ anymore.}
%We compose a sentence pair following a Wikipedia category page, i.e., given a $c$ as the first sentence, the second sentence is one of its candidate parent category $c_p$ returned by parent category identification.}
Importantly, the pre-trained BERT model has a potential data leakage issue, as the category we want to generate may exist in the corpus that was used for training.  Therefore, BERT may have an unfair advantage.  Nevertheless, it can be meaningful to see how our approaches fare against it.

\section{Evaluation}
\label{sec:eva}

%
%\begin{figure}[t]
%\centering
%   \includegraphics[width=0.5\textwidth, left]{figures/pci_recall.png} 
%   \caption{Impact of the cutoff parameter \emph{k} when finding topic terms for parent category identification.}
%\label{fig:pci_recall}
%\end{figure}

% -------SIG TESTING RESULTS (p value)-------
%pg VS NATS:  1.1072862970754967e-16
%pg VS NATS:  0.00743580900393899
%pg VS NATS:  0.1290410526189876
%pg VS NATS:  6.775955309503759e-09
%fast VS NATS:  7.415387531658004e-23
%fast VS NATS:  3.546387815930418e-24
%fast VS NATS:  1.7237445575742147e-26
%fast VS NATS:  8.869531823039424e-24
%abc VS fast:  8.414245751378347e-27
%abc VS fast:  1.689520425775575e-165
%abc VS fast:  1.6931388699492085e-134
%abc VS fast:  1.7343293075566934e-105
% ----------------------------------------
\begin{table}[t]
  \centering
% \vspace*{-0.25\baselineskip}  
  \caption{Candidate generation results using three generators (row 1-3) and our initial ranker (row 4). Statistical significance for row $i>1$ is tested against row $i-1$.
%  is tested against in a sequential pairwise manner, i.e. (A) vs. (B), (B) vs. (C), and (C) vs. (A)+(B)+(C).  
}
% \vspace*{-0.5\baselineskip}
  \begin{tabular}{lllll}
    \toprule
    \textbf{Method}  & \textbf{NDCG@1} & \textbf{NDCG@10} & \textbf{P@5} & \textbf{MRR@10}   \\
    \midrule
	PG~\citep{See:2017:GTP} & 0.1213 & 0.2151&0.0614&0.1794 \\% & 0.3101 & 0.2030  \\
	NATS~\citep{Shi:2018:NAT} & 0.2351$^\ddag$ & 0.2459$^\ddag$&0.0573&0.2483$^\ddag$\\%& 0.4850 & 0.3446  \\
	FAST~\citep{Chen:2018:FAS} & \textbf{0.3735$^\ddag$} & \textbf{0.3821$^\ddag$}&\textbf{0.0948$^\ddag$}&\textbf{0.3876$^\ddag$}\\%& \textbf{0.5478} & \textbf{0.4326}  \\
    \midrule
    SCG (Ours) & \textbf{0.5137$^\ddag$} & \textbf{0.8098$^\ddag$}&\textbf{0.2383$^\ddag$}&\textbf{0.6759$^\ddag$}\\%& \textbf{0.5927} & \textbf{0.5419}  \\  
    \bottomrule
  \end{tabular}
  \label{tbl:cg}
% \vspace*{-0.5\baselineskip}
\end{table}

\begin{table*}[t]
  \centering   
% +MRR
% TOP 3 BLOKS - KEEP 1000
% DROP 	last 2
% k or greedy
  \caption{Parent category identification results. Statistical significance of the bottom block is tested against the top block ($^\dag$/$^\ddag$) and columns two and three against column one (using $^{\scriptscriptstyle\lozenge}$/$^{\scriptscriptstyle\blacklozenge}$ to denote significance at the 0.05 and 0.01 level, respectively).}
%   \vspace*{-0.5\baselineskip}
  \begin{tabular}{lllllllll}
    \toprule
    \textbf{Method} & \multicolumn{2}{c}{\textbf{BM25}} && \multicolumn{2}{c}{\textbf{Hierarchy-based}} && \multicolumn{2}{c}{\textbf{Combined}} \\
    \cline{2-3}  \cline{5-6} \cline{8-9}
      & \textbf{MAP@k} & \textbf{MRR@k} && \textbf{MAP@k} & \textbf{MRR@k} && \textbf{MAP@k} & \textbf{MRR@k}  \\
    
    \midrule
	 Without query expansion ($k=10$)  & 0.0714 & 0.1423 && 0.1110$^{\scriptscriptstyle\blacklozenge}$ & 0.2342$^{\scriptscriptstyle\blacklozenge}$ && 0.1115 & 0.2348 \\
	 Without query expansion ($k=1000$)  & 0.0839 &0.1536&& 0.2153$^{\ddag\scriptscriptstyle\blacklozenge}$ &0.3639$^{\ddag\scriptscriptstyle\blacklozenge}$ && \textbf{0.2627}$^{\ddag\scriptscriptstyle\blacklozenge}$ & \textbf{0.4476}$^{\ddag\scriptscriptstyle\blacklozenge}$ \\
%	\midrule
%	(B) \citet{Lawrie:2001:FTW}+BM25 (k=2,top 10) & 0.3474 & 0.1736$\ddag$ &0.3133$\ddag$&& 0.1863$\ddag$ &0.3351$\ddag$&& 0.1965$\ddag$& 0.3539$\ddag$ \\
%	(B) \citet{Lawrie:2001:FTW}+BM25 (k=2,top 1000) & \textbf{0.7045} & 0.1884$\ddag$ &0.3212$\ddag$&& 0.1195 &0.2247&& 0.1783 & 0.3195 \\
	\midrule % thres=0.5
	 With query expansion ($k=10$)  & 0.1901$^\ddag$ &0.3347$^\ddag$&& 0.1920$^\ddag$ &0.3302$^\ddag$&& 0.2015$^\ddag$ &0.3492$^\ddag$ \\
	 With query expansion ($k=1000$) & \textbf{0.2036$^\ddag$} & \textbf{0.3414$^\ddag$} && 0.1160 &0.2210&& 0.1777 & 0.3178 \\
    \bottomrule
  \end{tabular}
  \label{tbl:cgi}
\end{table*}
\begin{table}[t]
  \centering
  % only keep the features ..(exclude subcat...)
  % same metrics with table 5
  \caption{Final ranking (end-to-end category generation) results. %The metrics are reported based on the same data set from candidate generation (top10 researved). 
  Statistical significance for lines 3 and 4 is tested against line 2, and for line 5 it is tested against line 3.}
% \vspace*{-0.5\baselineskip}  
%  \kb{Report only on Greedy (but in the text it should be called dynamic thresholding)}}
  \begin{tabular}{ll@{~~}l@{~~}llll}
    \toprule
    \textbf{Method} & \textbf{NDCG@1} & \textbf{NDCG@10}& \textbf{P@5} & \textbf{MRR@10} \\%& \textbf{ROUGE-1} & \textbf{ROUGE-2}  \\%& \# cand \\
    \midrule
    BERT~\citep{Devlin:2018:BPD} & 0.5308 & 0.8261& \textbf{0.2497} &0.6937\\%& \textbf{0.6083} & \textbf{0.5513}  \\%& 3.4 \\
    %  0.5336, 0.8401, 0.2530, 0.7009
    \midrule
     Features I & 0.5156 & 0.8149 & 0.2402 & 0.6779 \\%& 0.5945 & 0.5419 \\
     Features I+II  & 0.5516$^\dag$ & 0.8290 &0.2464$^\dag$&0.7035$^\dag$\\%& 0.5930 & 0.5422  \\%& 3.3 \\% 3.3
     Features I+III  & 0.5744$^\ddag$  & 0.8372$^\ddag$&0.2421&0.7195$^\ddag$\\%& 0.5941 & 0.5425   \\%& 3.8 \\ % 0.5 thres
    \midrule
    Features I+II+III  & \textbf{0.6028$^\ddag$} & \textbf{0.8423$^\dag$}&0.2445& \textbf{0.7363$^\ddag$} \\%& 0.5930 & 0.5420  \\
    % 0.5924, 0.8423, 0.2453, 0.7314
    \bottomrule
  \end{tabular}
  \label{tbl:fr}
% \vspace*{-0.5\baselineskip}  
\end{table}

We evaluate the performance of category generation (Sect.~\ref{sec:eva:cg}), parent category identification (Sect.~\ref{sec:eva:pci}), and final ranking (Sect.~\ref{sec:eva:fr}).

\if 0
As a writer, I usually put each experiment in its own subsection.  This makes it easier for reviewers to see the number of experiments / research questions that are addressed.  It also forces me to be careful about how each experiment is presented, i.e., each experiment should i) describe its purpose, ii) describe how it was done (if necessary), iii) analyze results, and iv) possibly draw conclusions about the research question or say what was learned from the experiment (sometimes iii and iv are combined).
\fi

\subsection{Category Generation}
\label{sec:eva:cg}

The first component of our pipeline is responsible for the generation of specific candidates that are relevant to the input entity set.  The results of the three generation models are presented in the top block of Table~\ref{tbl:cg}.
Pointer-generator (PG) generates 3.5 candidate categories on average for entity set, while NATS and FAST produce 1.5 and 1.6 candidates, respectively. 
These are ranked by the generation confidence score.
After pooling these together, each entity set yields 4.5 candidates on average.
Comparing the effectiveness of the three approaches, we can see that NATS outperforms PG significantly, thanks to its additional features such as the coverage mechanism. 
FAST outperforms both PG and NATS substantially and significantly. 
This tells us that making use of the salient information (in this case: tokens scattered in the entity set) from the extraction process can improve the performance of seq2seq models 
for category generation.

Our initial ranker, \emph{SCG} (short for Simple Category Generator), then ranks these candidates based on structure- and content-based features. The results are displayed in the last row of Table~\ref{tbl:cg}.
We find that our inexpensive features are very effective in sorting the candidates, improving all metrics substantially and statistically significantly. 
It should be noted that the comparison between SCG and the individual generators (PG, NATS, and FAST) is not a fair one, as SCF has access to all candidates produced by the individual generators. Nevertheless, these results show that the abstractive summarization methods can produce complementary results to each other, and our SCG method can effectively rank these candidates.

%\vspace*{-0.5\baselineskip}
\subsection{Parent Category Identification}
\label{sec:eva:pci}

Next, we evaluate the capability of our parent category identification approach against the test collection developed for this subtask (cf. Sect.~\ref{sec:data:tc:pgi}).  The results are reported in Table~\ref{tbl:cgi}, using two settings: considering top 10 or top 1000 candidates returned by BM25, and then re-ranking them.  We expect that going deeper in the initial BM25 ranking helps to increase coverage, but it also makes the ranking task more difficult by introducing noise.
We consider two ways to compose the keyword query.
First, we take the name of the category $c$ as is (top block).
Second, we apply query expansion and use the expanded query $\tilde{c}$ (bottom block). % by setting a dynamic threshold to extend the $c$  
The columns correspond to the retrieval method that is used: BM25 (first column), hierarchy-based (second column), and the combined method (last column).
 
The first observation is that our query expansion method substantially improves effectiveness (rows 1 vs. 3 and 2 vs. 4 in the first column). 
This shows that query expansion could narrow the vocabulary gap caused by the lack of shared tokens between the category and its parent.

Concerning the comparison of the three ranking models without query expansion (first block of three columns), 
hierarchy-based outperforms BM25 significantly, and the combined method achieves further substantial and significant improvements for $k=1000$.  Notably, for $k=10$, we do not observe the same improvements.  
This attests to the utility of the topic graph for effectively ranking a large number of candidates.

When combining the query expansion and ranking methods (bottom block and third column), the performance varies against the settings.
With $k=10$, the method utilizing the expanded query and combined method performs the best; this is the setting that we use for the final ranking, given its good trade-off between effectiveness and computational efficiency. 
Overall, the combined method without query expansion with $k=1000$ performs best. 
These results indicate that it is better to find the topic words in the late phase instead of refining the query at the beginning when we have many candidates. Otherwise, when the set of candidates is small, it is best to combine query expansion with the hierarchy-based method.
In summary, the best strategy to identify the parent categories is to perform query refinement and choose the ranking method accordingly, depending on the size of the candidate set.

\begin{table*}[!th]
  \centering
  \caption{Example showing the steps of category generation for a given input entity set.}
%\vspace*{-0.75\baselineskip}
\footnotesize
  \begin{tabular}{llp{12cm}}
    \toprule
    \emph{\textbf{Input}} &
    \textbf{List of entities} & ... Falkland Islands, Gibraltar, Gotland,  Greenland, Guernsey, Hitra, Isle of Wight, Jersey, Rhodes, Saare County...\\
    % \midrule
    & \textbf{Context} &  ...The Isle of Man are not members of FIFA or UEFA, as the Isle of Man FA are members of The Football Association (The FA), with similar status to an English county. Since they are not a member of either FIFA or UEFA, they are not eligible to enter either the World Cup or European Championship.... \\
     \midrule
     \multicolumn{2}{l}{\emph{\textbf{Ground truth}}} & European national and official selection-teams not affiliated to FIFA
     \\
     \midrule
     \emph{\textbf{Category generation}} & \textbf{SCG} & National football teams in the isle of man, Football teams in the isle of man, Isle of man, European national and official selection-teams not affiliated to FIFA \\ 
     \midrule
     \emph{\textbf{Category selection}} &  \textbf{BERT} & 1. National football teams in the isle of man \\
     & &  2. Football teams in the isle of man \\
     & & 3. Isle of man \\
     & & 4. \emph{European national and official selection-teams not affiliated to FIFA} \\
     \cline{2-3}
    & \textbf{Features I+II+III}  & 1. \emph{European national and official selection-teams not affiliated to FIFA} \\
    & & 2. Isle of man \\
    & & 3. Football teams in the isle of man \\
    & & 4. National football teams in the isle of man \\
    \bottomrule
  \end{tabular}
  \label{tbl:run-exam}
\end{table*}
%

%\vspace*{-0.25\baselineskip}
\subsection{Final Ranking}
\label{sec:eva:fr}

Given the candidate set returned by initial ranking, we identify the corresponding parent categories, and consider $\langle c,c_p \rangle$ pairs as input to the final ranking step.  Thus, the results we present in Table~\ref{tbl:fr} are to be seen as the end-to-end evaluation for the whole pipeline. (Note that this step also considers the parent categories, thus the results are not directly comparable to the initial ranking results we presented earlier.)
We take BERT as a baseline, and we report its results in the first line of Table~\ref{tbl:fr}.
For our proposed approach, we report on different combinations of features, such as only structure features (line 2), structure features with content features (line 3), structure features with importance features (line 4), and all features (last line).
Comparing the different types of features, we find that content-based features complement the structure features (line 3 vs. line 2), seen by the increase in scores on all the metrics. Importance features also enhance performance considerably (line 4 vs. line 2).
When combining all the features (line 5), we observe further performance improvements.
Importance features complement the rest of the features, as can be seen by the increase between line 3 and line 5.
As for the comparison against BERT, the NDCG@k and MRR results show that BERT is good at finding items, but not that good at ranking them. Our best method (line 5) outperforms BERT significantly on all the metrics except P@5.

\if 0 % >> to a fig instead
\begin{table*}[t]
  \centering
%   \footnotesize
  \caption{The distribution of the number of tables in the test set based on the reciprocal rank (RR) score. 
Acc \% signifies the accumulated percentage that the system can recommend at least one good category by recommending top $k$ categories. 
$k$ ranges from 1 to 10.
$Acc \% = \frac{\# tables'}{|T^{test}|}$ where $\# tables'$ is the accumulated number of tables and $T^{test}$ denotes the test set tables. There are 1055 in total (10\% of the whole training data).  
}
  \begin{tabular}{l|llllllllll}
    \toprule
    \toprule
    \textbf{RR}  & 1 & 1/2 & 1/3 & 1/4 & 1/5 & 1/6 & 1/7 & 1/8 & 1/9 & 1/10  \\
    \midrule
     \textbf{\# tables} & 636 & 152 & 84 & 68 & 51 & 36 & 14 & 8 & 4 & 2 \\  
     \textbf{Acc \%} & 60.2 & 74.7 & 82.7 & 89.1 & 93.9 & 97.3 & 98.7 & 99.4 & 99.8 & 1 \\
    \bottomrule
  \end{tabular}
  \label{tbl:rr}
\end{table*}
%
\fi

\section{Analysis}

We perform additional analysis of our results in this section. 
First, we present a running example, shown in Table~\ref{tbl:run-exam}, to illustrate each of the steps of our pipeline (cf. Sect.~\ref{sec:ps}) for generating and ranking new categories. 
In this example, the input set of entities is 14 teams in the Isle of Man that are not affiliated to FIFA.
The ground truth category, as well as its parent categories, are removed from training data.
Apart from the ground truth category, our SCG candidate generator produces three additional candidate categories; these four candidates are then ranked by BERT and by our method (using all feature sets).  The respective rankings are shown in the bottom two lines in Table~\ref{tbl:run-exam}.
Comparing with BERT, which put the ground truth category in the last place, our method successfully placed it at the top rank.

\begin{figure}[t]
   \centering
    \vspace*{-1\baselineskip}
   \includegraphics[width=0.45\textwidth]{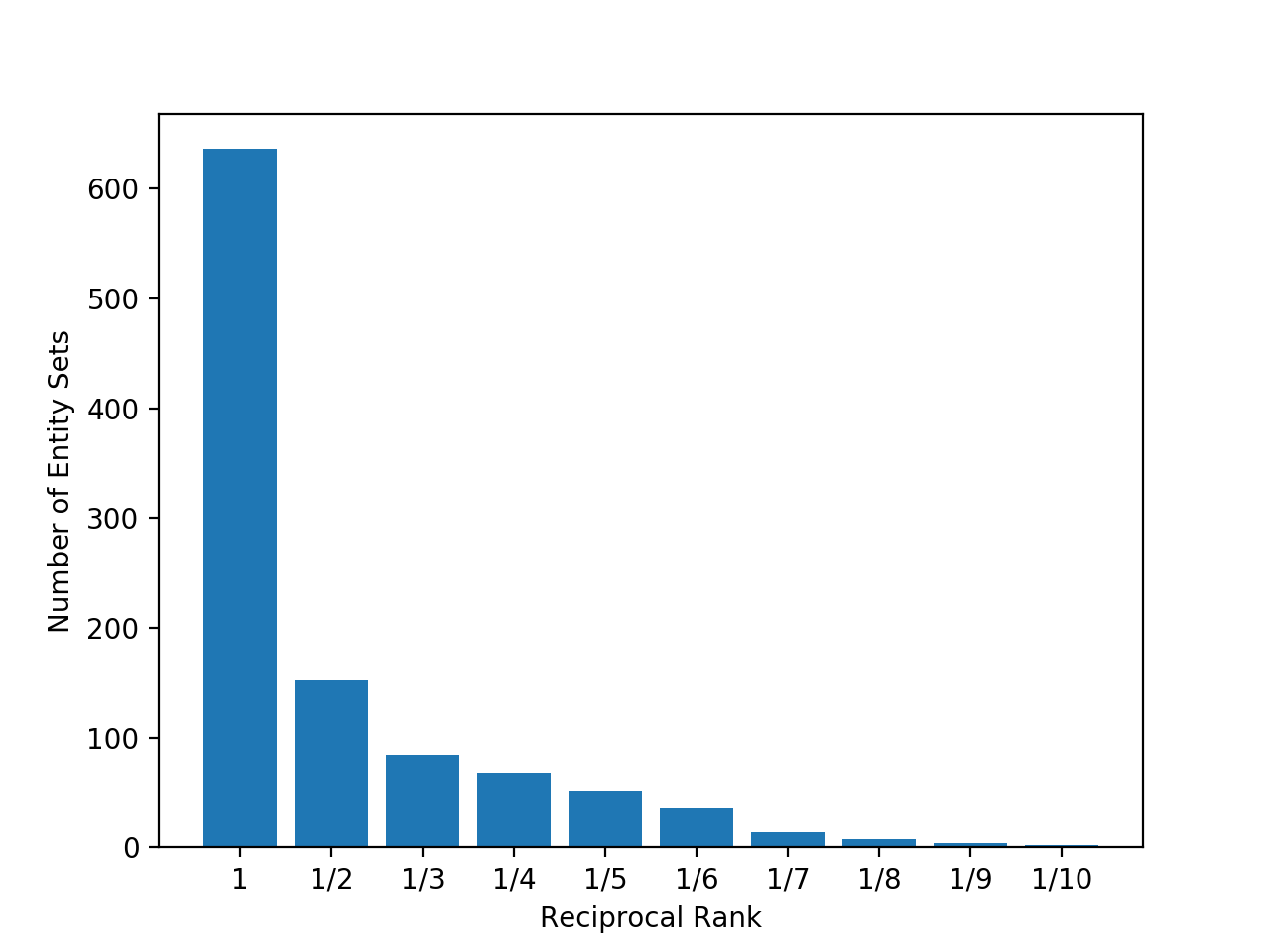} 
    \vspace*{-0.75\baselineskip}
   \caption{The distribution of the number of entity sets in the test set based on the reciprocal rank at which the first relevant category suggestion is returned.} 
   \vspace*{-0.5\baselineskip}
% The number signifies the cases where we can recommend at least one good category in by recommending top $k$ categories. 
% $k$ ranges from 1 to 10.
% There are 1055 in total (10\% of the whole training data).}
\label{fig:rr}
\end{figure}

To investigate the practical utility of our approach, we measure how often it can return a good recommendation at a high rank position.  Figure~\ref{fig:rr} shows the distribution of all test cases (i.e., input entity sets) with respect to reciprocal rank.  That is, if the first relevant category suggestion was returned at rank position $k$, then the reciprocal rank is $1/k$. 
We find that in over 60\% of all test cases, a relevant suggestion is returned at the top rank.
Further, in over 94\% of the input cases there is always at least one relevant suggestion returned within the top 5 rank positions.
These results demonstrate that our approach has great merit to be deployed in a practical environment, e.g., as a category suggestion service in Wikipedia.

% how often the system can suggest a good category by the top recommendation, we \sz{demonstrate} the number of entity sets based on the reciprocal rank score using the ranking results of the combined method.
% In terms of $Acc \%$, it can recommend relevant categories by top-1 suggestions for 60.2\% of the test entity sets.
% %
% % Alternative analysis
% To verify the practical utility of our system, we check how often it can recommend a relevant category within the top 5 suggestions.  We find that in over 94\% of the input cases there is always at least one relevant suggestion and it happens in less than 6\% of the time that no relevant suggestion is returned in the top 5.
%
%In conclusion, our approach recommends categories that satisfy the specificity, hierarchy, redundancy and importance requirements.

%

\vspace*{-0.25\baselineskip}
\section{Conclusion}

Category systems of large-scale knowledge bases are unique and valuable resources that can be utilized in a wide range of information access tasks.  However, they are currently created and maintained manually by editors. 
This paper presents a pipeline approach to generate categories in an automatic manner, given a set of entities and their context as input.  %One particular challenge involved here is finding a suitable name for the category.  Based on Wikipedia's category labeling guidelines, 
We identify four challenges of automatic category generation, which are specificity, hierarchy, redundancy, and importance. 
%To overcome these challenges, we propose a pipeline with several steps, which include candidate generation, initial ranking, parent category identification, and the final ranking, respectively.

To address the challenge of \emph{specificity}, the first task of candidate generation aims to generate candidate categories that are specific enough given the entity set and context. 
%For the lack of existing test collections, we conduct a strategy sampling to construct the training data.
Given an entity set, we use abstractive summarization models to generate candidates. 
The candidates are initially ranked, leveraging a set of inexpensive features, based on structure and content, to prune the candidate set.
Experimental results show that the abstractive summarization models can generate specific candidates, and the initial ranker can select the most suitable ones.

Parent category identification aims at addressing \emph{hierarchy} and \emph{redundancy} by finding the location of the candidate category in the current category system.
To fill the gap caused by the insufficiency of term-based matching and (possible) violation of the transitivity principle, we build a topic graph leveraging all category-parent pairs in the category system. It is utilized in two ways, to expand the query and to rank the parent candidates. Experimental results show that the topic graph can enhance performance by either query expansion or by hierarchy-based ranking.

The final ranking step aims to address the challenge of \emph{importance} by ranking category-parent pairs. 
Apart from structure- and content-based features, it also considers importance features, which tap into the characteristics of member entities to approximate category importance.  
% such as the category or member entity size, can reflect the importance
This effectively complements the other two groups of features. 

We develop a test collection based on Wikipedia categories and perform both end-to-end and component-level evaluation.  We show the effectiveness of our approach against a BERT-based baseline and also demonstrate that performance is strong enough to be deployed in a practical application.

\bibliographystyle{ACM-Reference-Format}
\bibliography{00paper}


\begin{thebibliography}{43}

%%% the Web address.

\ifx \showCODEN    \undefined \def \showCODEN     #1{\unskip}     \fi
\ifx \showDOI      \undefined \def \showDOI       #1{#1}\fi
\ifx \showISBNx    \undefined \def \showISBNx     #1{\unskip}     \fi
\ifx \showISBNxiii \undefined \def \showISBNxiii  #1{\unskip}     \fi
\ifx \showISSN     \undefined \def \showISSN      #1{\unskip}     \fi
\ifx \showLCCN     \undefined \def \showLCCN      #1{\unskip}     \fi
\ifx \shownote     \undefined \def \shownote      #1{#1}          \fi
\ifx \showarticletitle \undefined \def \showarticletitle #1{#1}   \fi
\ifx \showURL      \undefined \def \showURL       {\relax}        \fi
% invisible to TeX
\providecommand\bibfield[2]{#2}
\providecommand\bibinfo[2]{#2}
\providecommand\natexlab[1]{#1}
\providecommand\showeprint[2][]{arXiv:#2}

\bibitem[\protect\citeauthoryear{Al-Arfaj and Al-Salman}{Al-Arfaj and
  Al-Salman}{2015}]%
        {Abeer:2015:OCT}
\bibfield{author}{\bibinfo{person}{Abeer Al-Arfaj} {and}
  \bibinfo{person}{AbdulMalik Al-Salman}.} \bibinfo{year}{2015}\natexlab{}.
\newblock \showarticletitle{Ontology Construction from Text: Challenges and
  Trends}.
\newblock \bibinfo{journal}{\emph{International Journal of Artificial
  Intelligence and Expert Systems}}  \bibinfo{volume}{6}, Article
  \bibinfo{articleno}{2} (\bibinfo{year}{2015}),
  \bibinfo{numpages}{15-26}~pages.
\newblock


\bibitem[\protect\citeauthoryear{Balog, Bron, and De~Rijke}{Balog
  et~al\mbox{.}}{2011}]%
        {Balog:2011:QME}
\bibfield{author}{\bibinfo{person}{Krisztian Balog}, \bibinfo{person}{Marc
  Bron}, {and} \bibinfo{person}{Maarten De~Rijke}.}
  \bibinfo{year}{2011}\natexlab{}.
\newblock \showarticletitle{Query modeling for entity search based on terms,
  categories, and examples}.
\newblock \bibinfo{journal}{\emph{ACM Trans. Inf. Syst.}} \bibinfo{volume}{29},
  \bibinfo{number}{4}, Article \bibinfo{articleno}{22} (\bibinfo{date}{Dec.}
  \bibinfo{year}{2011}), \bibinfo{numpages}{22:1--22:31}~pages.
\newblock


\bibitem[\protect\citeauthoryear{Bedini}{Bedini}{2007}]%
        {Bedini:2007:AOG}
\bibfield{author}{\bibinfo{person}{Ivan Bedini}.}
  \bibinfo{year}{2007}\natexlab{}.
\newblock \showarticletitle{Automatic Ontology Generation : State of the Art}.
\newblock \bibinfo{journal}{\emph{Molecular Evolution}}  \bibinfo{volume}{44},
  Article \bibinfo{articleno}{2} (\bibinfo{year}{2007}),
  \bibinfo{numpages}{226-233}~pages.
\newblock


\bibitem[\protect\citeauthoryear{Biemann}{Biemann}{2005}]%
        {Biemann:2005:OLT}
\bibfield{author}{\bibinfo{person}{Chris Biemann}.}
  \bibinfo{year}{2005}\natexlab{}.
\newblock \showarticletitle{Ontology Learning from Text: {A} Survey of
  Methods}.
\newblock \bibinfo{journal}{\emph{LDV Forum}} \bibinfo{volume}{20},
  \bibinfo{number}{2} (\bibinfo{year}{2005}), \bibinfo{pages}{75--93}.
\newblock


\bibitem[\protect\citeauthoryear{Blei, Griffiths, and Jordan}{Blei
  et~al\mbox{.}}{2010}]%
        {Blei:2010:NCR}
\bibfield{author}{\bibinfo{person}{David~M. Blei}, \bibinfo{person}{Thomas~L.
  Griffiths}, {and} \bibinfo{person}{Michael~I. Jordan}.}
  \bibinfo{year}{2010}\natexlab{}.
\newblock \showarticletitle{The Nested Chinese Restaurant Process and Bayesian
  Nonparametric Inference of Topic Hierarchies}.
\newblock \bibinfo{journal}{\emph{J. ACM}} \bibinfo{volume}{57},
  \bibinfo{number}{2}, Article \bibinfo{articleno}{7} (\bibinfo{date}{Feb.}
  \bibinfo{year}{2010}), \bibinfo{numpages}{30}~pages.
\newblock
\showISSN{0004-5411}


\bibitem[\protect\citeauthoryear{Boldi and Monti}{Boldi and Monti}{2016}]%
        {Boldi:2016:CWC}
\bibfield{author}{\bibinfo{person}{Paolo Boldi} {and} \bibinfo{person}{Corrado
  Monti}.} \bibinfo{year}{2016}\natexlab{}.
\newblock \showarticletitle{Cleansing Wikipedia Categories Using Centrality}.
  In \bibinfo{booktitle}{\emph{Proceedings of the 25th International Conference
  Companion on World Wide Web}} \emph{(\bibinfo{series}{WWW '16 Companion})}.
  \bibinfo{pages}{969--974}.
\newblock
\showISBNx{978-1-4503-4144-8}


\bibitem[\protect\citeauthoryear{Chen and Bansal}{Chen and Bansal}{2018}]%
        {Chen:2018:FAS}
\bibfield{author}{\bibinfo{person}{Yen-Chun Chen} {and} \bibinfo{person}{Mohit
  Bansal}.} \bibinfo{year}{2018}\natexlab{}.
\newblock \showarticletitle{Fast Abstractive Summarization with
  Reinforce-Selected Sentence Rewriting}. In
  \bibinfo{booktitle}{\emph{Proceedings of the 56th Annual Meeting of the
  Association for Computational Linguistics (Volume 1: Long Papers)}}.
  \bibinfo{pages}{675--686}.
\newblock


\bibitem[\protect\citeauthoryear{Choi, Levy, Choi, and Zettlemoyer}{Choi
  et~al\mbox{.}}{2018}]%
        {Choi:2018:UET}
\bibfield{author}{\bibinfo{person}{Eunsol Choi}, \bibinfo{person}{Omer Levy},
  \bibinfo{person}{Yejin Choi}, {and} \bibinfo{person}{Luke Zettlemoyer}.}
  \bibinfo{year}{2018}\natexlab{}.
\newblock \showarticletitle{Ultra-Fine Entity Typing}. In
  \bibinfo{booktitle}{\emph{Proceedings of the 56th Annual Meeting of the
  Association for Computational Linguistics}}. \bibinfo{pages}{87--96}.
\newblock


\bibitem[\protect\citeauthoryear{Ciglan, N{\o}rv{\aa}g, and Hluch\'{y}}{Ciglan
  et~al\mbox{.}}{2012}]%
        {Ciglan:2012:SMA}
\bibfield{author}{\bibinfo{person}{Marek Ciglan}, \bibinfo{person}{Kjetil
  N{\o}rv{\aa}g}, {and} \bibinfo{person}{Ladislav Hluch\'{y}}.}
  \bibinfo{year}{2012}\natexlab{}.
\newblock \showarticletitle{The SemSets Model for Ad-hoc Semantic List Search}.
  In \bibinfo{booktitle}{\emph{Proceedings of the 21st International Conference
  on World Wide Web}} \emph{(\bibinfo{series}{WWW '12})}.
  \bibinfo{pages}{131--140}.
\newblock
\showISBNx{978-1-4503-1229-5}


\bibitem[\protect\citeauthoryear{Devlin, Chang, Lee, and Toutanova}{Devlin
  et~al\mbox{.}}{2018}]%
        {Devlin:2018:BPD}
\bibfield{author}{\bibinfo{person}{Jacob Devlin}, \bibinfo{person}{Ming{-}Wei
  Chang}, \bibinfo{person}{Kenton Lee}, {and} \bibinfo{person}{Kristina
  Toutanova}.} \bibinfo{year}{2018}\natexlab{}.
\newblock \showarticletitle{{BERT:} Pre-training of Deep Bidirectional
  Transformers for Language Understanding}.
\newblock \bibinfo{journal}{\emph{CoRR}}  \bibinfo{volume}{abs/1810.04805}
  (\bibinfo{year}{2018}).
\newblock


\bibitem[\protect\citeauthoryear{Graus, Odijk, and de~Rijke}{Graus
  et~al\mbox{.}}{2018}]%
        {Graus:2018:BCM}
\bibfield{author}{\bibinfo{person}{David Graus}, \bibinfo{person}{Daan Odijk},
  {and} \bibinfo{person}{Maarten de Rijke}.} \bibinfo{year}{2018}\natexlab{}.
\newblock \showarticletitle{The Birth of Collective Memories: Analyzing
  Emerging Entities in Text Streams}.
\newblock \bibinfo{journal}{\emph{Journal of the Association for Information
  Science and Technology}} \bibinfo{volume}{69}, \bibinfo{number}{6}
  (\bibinfo{year}{2018}), \bibinfo{pages}{773--786}.
\newblock


\bibitem[\protect\citeauthoryear{Hancock, Lee, and Yu}{Hancock
  et~al\mbox{.}}{2019}]%
        {Braden:2019:GTW}
\bibfield{author}{\bibinfo{person}{Braden Hancock}, \bibinfo{person}{Hongrae
  Lee}, {and} \bibinfo{person}{Cong Yu}.} \bibinfo{year}{2019}\natexlab{}.
\newblock \showarticletitle{Generating Titles for Web Tables}. In
  \bibinfo{booktitle}{\emph{The World Wide Web Conference}}
  \emph{(\bibinfo{series}{WWW '19})}. \bibinfo{pages}{638--647}.
\newblock
\showISBNx{978-1-4503-6674-8}


\bibitem[\protect\citeauthoryear{Hoffart, Altun, and Weikum}{Hoffart
  et~al\mbox{.}}{2014}]%
        {Hoffart:2014:DEE}
\bibfield{author}{\bibinfo{person}{Johannes Hoffart}, \bibinfo{person}{Yasemin
  Altun}, {and} \bibinfo{person}{Gerhard Weikum}.}
  \bibinfo{year}{2014}\natexlab{}.
\newblock \showarticletitle{Discovering Emerging Entities with Ambiguous
  Names}. In \bibinfo{booktitle}{\emph{Proceedings of the 23rd International
  Conference on World Wide Web}} \emph{(\bibinfo{series}{WWW '14})}.
  \bibinfo{pages}{385--396}.
\newblock


\bibitem[\protect\citeauthoryear{Ji and Grishman}{Ji and Grishman}{2011}]%
        {Ji:2011:KBP}
\bibfield{author}{\bibinfo{person}{Heng Ji} {and} \bibinfo{person}{Ralph
  Grishman}.} \bibinfo{year}{2011}\natexlab{}.
\newblock \showarticletitle{Knowledge Base Population: Successful Approaches
  and Challenges}. In \bibinfo{booktitle}{\emph{Proceedings of the 49th Annual
  Meeting of the Association for Computational Linguistics: Human Language
  Technologies - Volume 1}} \emph{(\bibinfo{series}{HLT '11})}.
  \bibinfo{pages}{1148--1158}.
\newblock


\bibitem[\protect\citeauthoryear{Kaptein and Kamps}{Kaptein and Kamps}{2013}]%
        {Kaptein:2013:ECS}
\bibfield{author}{\bibinfo{person}{Rianne Kaptein} {and} \bibinfo{person}{Jaap
  Kamps}.} \bibinfo{year}{2013}\natexlab{}.
\newblock \showarticletitle{Exploiting the Category Structure of Wikipedia for
  Entity Ranking}.
\newblock \bibinfo{journal}{\emph{Artif. Intell.}}  \bibinfo{volume}{194}
  (\bibinfo{date}{Jan.} \bibinfo{year}{2013}), \bibinfo{pages}{111--129}.
\newblock
\showISSN{0004-3702}


\bibitem[\protect\citeauthoryear{Kirillovich and Nevzorova}{Kirillovich and
  Nevzorova}{2018}]%
        {Kirillovich:2018:OAO}
\bibfield{author}{\bibinfo{person}{Alexander Kirillovich} {and}
  \bibinfo{person}{Olga Nevzorova}.} \bibinfo{year}{2018}\natexlab{}.
\newblock \showarticletitle{Ontological Analysis of the Wikipedia Category
  System}. In \bibinfo{booktitle}{\emph{Proceedings of the 10th International
  Joint Conference on Knowledge Discovery, Knowledge Engineering and Knowledge
  Management, (IC3K '18)}}. \bibinfo{pages}{356--364}.
\newblock


\bibitem[\protect\citeauthoryear{Lauer}{Lauer}{1996}]%
        {Lauer:1996:DSL}
\bibfield{author}{\bibinfo{person}{Mark Lauer}.}
  \bibinfo{year}{1996}\natexlab{}.
\newblock \showarticletitle{Designing Statistical Language Learners:
  Experiments on Noun Compounds}.
\newblock \bibinfo{journal}{\emph{CoRR}}  \bibinfo{volume}{cmp-lg/9609008}
  (\bibinfo{year}{1996}).
\newblock


\bibitem[\protect\citeauthoryear{Lawrie, Croft, and Rosenberg}{Lawrie
  et~al\mbox{.}}{2001}]%
        {Lawrie:2001:FTW}
\bibfield{author}{\bibinfo{person}{Dawn Lawrie}, \bibinfo{person}{W.~Bruce
  Croft}, {and} \bibinfo{person}{Arnold Rosenberg}.}
  \bibinfo{year}{2001}\natexlab{}.
\newblock \showarticletitle{Finding Topic Words for Hierarchical
  Summarization}. In \bibinfo{booktitle}{\emph{Proceedings of the 24th Annual
  International ACM SIGIR Conference on Research and Development in Information
  Retrieval}} \emph{(\bibinfo{series}{SIGIR '01})}. \bibinfo{pages}{349--357}.
\newblock
\showISBNx{1-58113-331-6}


\bibitem[\protect\citeauthoryear{Lehmberg, Ritze, Meusel, and Bizer}{Lehmberg
  et~al\mbox{.}}{2016}]%
        {Lehmberg:2016:LPC}
\bibfield{author}{\bibinfo{person}{Oliver Lehmberg}, \bibinfo{person}{Dominique
  Ritze}, \bibinfo{person}{Robert Meusel}, {and} \bibinfo{person}{Christian
  Bizer}.} \bibinfo{year}{2016}\natexlab{}.
\newblock \showarticletitle{A Large Public Corpus of Web Tables Containing Time
  and Context Metadata}. In \bibinfo{booktitle}{\emph{Proceedings of the 25th
  International Conference Companion on World Wide Web}}
  \emph{(\bibinfo{series}{WWW '16 Companion})}. \bibinfo{pages}{75--76}.
\newblock
\showISBNx{978-1-4503-4144-8}


\bibitem[\protect\citeauthoryear{Ma, Chen, Chang, Du, Xu, and Chang}{Ma
  et~al\mbox{.}}{2018}]%
        {Ma:2018:LFW}
\bibfield{author}{\bibinfo{person}{Denghao Ma}, \bibinfo{person}{Yueguo Chen},
  \bibinfo{person}{Kevin Chen-Chuan Chang}, \bibinfo{person}{Xiaoyong Du},
  \bibinfo{person}{Chuanfei Xu}, {and} \bibinfo{person}{Yi Chang}.}
  \bibinfo{year}{2018}\natexlab{}.
\newblock \showarticletitle{Leveraging Fine-Grained Wikipedia Categories for
  Entity Search}. In \bibinfo{booktitle}{\emph{Proceedings of the 2018 World
  Wide Web Conference}} \emph{(\bibinfo{series}{WWW '18})}.
  \bibinfo{pages}{1623--1632}.
\newblock
\showISBNx{978-1-4503-5639-8}


\bibitem[\protect\citeauthoryear{McGuinness}{McGuinness}{2002}]%
        {McGuinness:2002:OCA}
\bibfield{author}{\bibinfo{person}{Deborah~L. McGuinness}.}
  \bibinfo{year}{2002}\natexlab{}.
\newblock \showarticletitle{Ontologies Come of Age}.
\newblock \bibinfo{journal}{\emph{Spinning the Semantic Web: Bringing the World
  Wide Web to Its Full Potential}} (\bibinfo{year}{2002}),
  \bibinfo{pages}{171--195}.
\newblock


\bibitem[\protect\citeauthoryear{Nadeau and Sekine}{Nadeau and Sekine}{2007}]%
        {Nadeau:2007:SNE}
\bibfield{author}{\bibinfo{person}{David Nadeau} {and} \bibinfo{person}{Satoshi
  Sekine}.} \bibinfo{year}{2007}\natexlab{}.
\newblock \showarticletitle{A survey of named entity recognition and
  classification}.
\newblock \bibinfo{journal}{\emph{Lingvisticae Investigationes}}
  \bibinfo{volume}{30}, \bibinfo{number}{1} (\bibinfo{year}{2007}),
  \bibinfo{pages}{3--26}.
\newblock


\bibitem[\protect\citeauthoryear{Nastase and Strube}{Nastase and
  Strube}{2008}]%
        {Nastase:2008:DWC}
\bibfield{author}{\bibinfo{person}{Vivi Nastase} {and} \bibinfo{person}{Michael
  Strube}.} \bibinfo{year}{2008}\natexlab{}.
\newblock \showarticletitle{Decoding Wikipedia Categories for Knowledge
  Acquisition}. In \bibinfo{booktitle}{\emph{Proceedings of the 23rd National
  Conference on Artificial Intelligence - Volume 2}}
  \emph{(\bibinfo{series}{AAAI'08})}. \bibinfo{pages}{1219--1224}.
\newblock
\showISBNx{978-1-57735-368-3}


\bibitem[\protect\citeauthoryear{Nastase and Strube}{Nastase and
  Strube}{2013}]%
        {Nastase:2013:TWL}
\bibfield{author}{\bibinfo{person}{Vivi Nastase} {and} \bibinfo{person}{Michael
  Strube}.} \bibinfo{year}{2013}\natexlab{}.
\newblock \showarticletitle{Transforming Wikipedia into a Large Scale
  Multilingual Concept Network}.
\newblock \bibinfo{journal}{\emph{Artif. Intell.}}  \bibinfo{volume}{194}
  (\bibinfo{date}{Jan.} \bibinfo{year}{2013}), \bibinfo{pages}{62--85}.
\newblock
\showISSN{0004-3702}


\bibitem[\protect\citeauthoryear{Obeidat, Fern, Shahbazi, and
  Tadepalli}{Obeidat et~al\mbox{.}}{2019}]%
        {Obeidat:2019:DZF}
\bibfield{author}{\bibinfo{person}{Rasha Obeidat}, \bibinfo{person}{Xiaoli
  Fern}, \bibinfo{person}{Hamed Shahbazi}, {and} \bibinfo{person}{Prasad
  Tadepalli}.} \bibinfo{year}{2019}\natexlab{}.
\newblock \showarticletitle{Description-Based Zero-shot Fine-Grained Entity
  Typing}. In \bibinfo{booktitle}{\emph{Proceedings of the 2019 Conference of
  the North {A}merican Chapter of the Association for Computational
  Linguistics: Human Language Technologies}}. \bibinfo{pages}{807--814}.
\newblock


\bibitem[\protect\citeauthoryear{Pa\c{s}ca}{Pa\c{s}ca}{2017}]%
        {Pasca:2017:GTV}
\bibfield{author}{\bibinfo{person}{Marius Pa\c{s}ca}.}
  \bibinfo{year}{2017}\natexlab{}.
\newblock \showarticletitle{German Typographers vs. German Grammar:
  Decomposition of Wikipedia Category Labels into Attribute-Value Pairs}. In
  \bibinfo{booktitle}{\emph{Proceedings of the Tenth ACM International
  Conference on Web Search and Data Mining}} \emph{(\bibinfo{series}{WSDM
  '17})}. \bibinfo{pages}{315--324}.
\newblock
\showISBNx{978-1-4503-4675-7}


\bibitem[\protect\citeauthoryear{Paulus, Xiong, and Socher}{Paulus
  et~al\mbox{.}}{2017}]%
        {Paulus:2017:DRM}
\bibfield{author}{\bibinfo{person}{Romain Paulus}, \bibinfo{person}{Caiming
  Xiong}, {and} \bibinfo{person}{Richard Socher}.}
  \bibinfo{year}{2017}\natexlab{}.
\newblock \showarticletitle{A Deep Reinforced Model for Abstractive
  Summarization}.
\newblock \bibinfo{journal}{\emph{CoRR}}  \bibinfo{volume}{abs/1705.04304}
  (\bibinfo{year}{2017}).
\newblock


\bibitem[\protect\citeauthoryear{Pivk}{Pivk}{2005}]%
        {Pivk:2005:AOG}
\bibfield{author}{\bibinfo{person}{Aleksander Pivk}.}
  \bibinfo{year}{2005}\natexlab{}.
\newblock \showarticletitle{Automatic Ontology Generation from Web Tabular
  Structures}.
\newblock \bibinfo{journal}{\emph{AI Communications}}  \bibinfo{volume}{19}
  (\bibinfo{year}{2005}), \bibinfo{pages}{83--85}.
\newblock


\bibitem[\protect\citeauthoryear{Ren, He, Qu, Huang, Ji, and Han}{Ren
  et~al\mbox{.}}{2016}]%
        {Ren:2016:AFE}
\bibfield{author}{\bibinfo{person}{Xiang Ren}, \bibinfo{person}{Wenqi He},
  \bibinfo{person}{Meng Qu}, \bibinfo{person}{Lifu Huang},
  \bibinfo{person}{Heng Ji}, {and} \bibinfo{person}{Jiawei Han}.}
  \bibinfo{year}{2016}\natexlab{}.
\newblock \showarticletitle{{AFET}: Automatic Fine-Grained Entity Typing by
  Hierarchical Partial-Label Embedding}. In
  \bibinfo{booktitle}{\emph{Proceedings of the 2016 Conference on Empirical
  Methods in Natural Language Processing}}.
\newblock


\bibitem[\protect\citeauthoryear{Ritze and Bizer}{Ritze and Bizer}{2017}]%
        {Ritze:2017:MWT}
\bibfield{author}{\bibinfo{person}{Dominique Ritze} {and}
  \bibinfo{person}{Christian Bizer}.} \bibinfo{year}{2017}\natexlab{}.
\newblock \showarticletitle{Matching Web Tables To DBpedia - A Feature Utility
  Study}. In \bibinfo{booktitle}{\emph{Proceedings of the 20th International
  Conference on Extending Database Technology (EDBT '17)}}.
  \bibinfo{pages}{210--221}.
\newblock


\bibitem[\protect\citeauthoryear{Ritze, Lehmberg, and Bizer}{Ritze
  et~al\mbox{.}}{2015}]%
        {Ritze:2015:MHT}
\bibfield{author}{\bibinfo{person}{Dominique Ritze}, \bibinfo{person}{Oliver
  Lehmberg}, {and} \bibinfo{person}{Christian Bizer}.}
  \bibinfo{year}{2015}\natexlab{}.
\newblock \showarticletitle{Matching HTML Tables to DBpedia}. In
  \bibinfo{booktitle}{\emph{Proceedings of the 5th International Conference on
  Web Intelligence, Mining and Semantics}} \emph{(\bibinfo{series}{WIMS '15})}.
  Article \bibinfo{articleno}{10}, \bibinfo{numpages}{6}~pages.
\newblock
\showISBNx{978-1-4503-3293-4}


\bibitem[\protect\citeauthoryear{Sanderson and Croft}{Sanderson and
  Croft}{1999}]%
        {Sanderson:1999:DCH}
\bibfield{author}{\bibinfo{person}{Mark Sanderson} {and} \bibinfo{person}{Bruce
  Croft}.} \bibinfo{year}{1999}\natexlab{}.
\newblock \showarticletitle{Deriving Concept Hierarchies from Text}. In
  \bibinfo{booktitle}{\emph{Proceedings of the 22Nd Annual International ACM
  SIGIR Conference on Research and Development in Information Retrieval}}
  \emph{(\bibinfo{series}{SIGIR '99})}. \bibinfo{pages}{206--213}.
\newblock
\showISBNx{1-58113-096-1}


\bibitem[\protect\citeauthoryear{See, Liu, and Manning}{See
  et~al\mbox{.}}{2017}]%
        {See:2017:GTP}
\bibfield{author}{\bibinfo{person}{Abigail See}, \bibinfo{person}{Peter~J.
  Liu}, {and} \bibinfo{person}{Christopher~D. Manning}.}
  \bibinfo{year}{2017}\natexlab{}.
\newblock \showarticletitle{Get To The Point: Summarization with
  Pointer-Generator Networks}. In \bibinfo{booktitle}{\emph{Proceedings of the
  55th Annual Meeting of the Association for Computational Linguistics (Volume
  1: Long Papers)}}. \bibinfo{pages}{1073--1083}.
\newblock


\bibitem[\protect\citeauthoryear{Shi, Keneshloo, Ramakrishnan, and Reddy}{Shi
  et~al\mbox{.}}{2018}]%
        {Shi:2018:NAT}
\bibfield{author}{\bibinfo{person}{Tian Shi}, \bibinfo{person}{Yaser
  Keneshloo}, \bibinfo{person}{Naren Ramakrishnan}, {and}
  \bibinfo{person}{Chandan~K. Reddy}.} \bibinfo{year}{2018}\natexlab{}.
\newblock \showarticletitle{Neural Abstractive Text Summarization with
  Sequence-to-Sequence Models}.
\newblock \bibinfo{journal}{\emph{CoRR}} (\bibinfo{year}{2018}).
\newblock


\bibitem[\protect\citeauthoryear{Stoica, Hearst, and Richardson}{Stoica
  et~al\mbox{.}}{2007}]%
        {Stoica:2007:ACH}
\bibfield{author}{\bibinfo{person}{Emilia Stoica}, \bibinfo{person}{Marti
  Hearst}, {and} \bibinfo{person}{Megan Richardson}.}
  \bibinfo{year}{2007}\natexlab{}.
\newblock \showarticletitle{Automating Creation of Hierarchical Faceted
  Metadata Structures}. In \bibinfo{booktitle}{\emph{Human Language
  Technologies 2007: The Conference of the North {A}merican Chapter of the
  Association for Computational Linguistics; Proceedings of the Main
  Conference}}. \bibinfo{pages}{244--251}.
\newblock


\bibitem[\protect\citeauthoryear{Sun, Xiao, Wang, and Wang}{Sun
  et~al\mbox{.}}{2015}]%
        {Sun:2015:OCL}
\bibfield{author}{\bibinfo{person}{Xiangyan Sun}, \bibinfo{person}{Yanghua
  Xiao}, \bibinfo{person}{Haixun Wang}, {and} \bibinfo{person}{Wei Wang}.}
  \bibinfo{year}{2015}\natexlab{}.
\newblock \showarticletitle{On Conceptual Labeling of a Bag of Words}. In
  \bibinfo{booktitle}{\emph{Proceedings of the 24th International Conference on
  Artificial Intelligence}} \emph{(\bibinfo{series}{IJCAI '15})}.
  \bibinfo{pages}{1326--1332}.
\newblock
\showISBNx{9781577357384}


\bibitem[\protect\citeauthoryear{Yosef, Bauer, Hoffart, Spaniol, and
  Weikum}{Yosef et~al\mbox{.}}{2012}]%
        {Yosef:2012:HHT}
\bibfield{author}{\bibinfo{person}{Mohamed~Amir Yosef}, \bibinfo{person}{Sandro
  Bauer}, \bibinfo{person}{Johannes Hoffart}, \bibinfo{person}{Marc Spaniol},
  {and} \bibinfo{person}{Gerhard Weikum}.} \bibinfo{year}{2012}\natexlab{}.
\newblock \showarticletitle{{HYENA}: Hierarchical Type Classification for
  Entity Names}. In \bibinfo{booktitle}{\emph{Proceedings of {COLING} 2012:
  Posters}}. \bibinfo{pages}{1361--1370}.
\newblock


\bibitem[\protect\citeauthoryear{Yu, Thom, and Tam}{Yu et~al\mbox{.}}{2007}]%
        {Yu:2007:OEU}
\bibfield{author}{\bibinfo{person}{Jonathan Yu}, \bibinfo{person}{James~A.
  Thom}, {and} \bibinfo{person}{Audrey Tam}.} \bibinfo{year}{2007}\natexlab{}.
\newblock \showarticletitle{Ontology Evaluation Using Wikipedia Categories for
  Browsing}. In \bibinfo{booktitle}{\emph{Proceedings of the Sixteenth ACM
  Conference on Conference on Information and Knowledge Management}}
  \emph{(\bibinfo{series}{CIKM '07})}. \bibinfo{pages}{223--232}.
\newblock
\showISBNx{978-1-59593-803-9}


\bibitem[\protect\citeauthoryear{Zhang and Balog}{Zhang and Balog}{2017}]%
        {Zhang:2017:ESA}
\bibfield{author}{\bibinfo{person}{Shuo Zhang} {and} \bibinfo{person}{Krisztian
  Balog}.} \bibinfo{year}{2017}\natexlab{}.
\newblock \showarticletitle{EntiTables: Smart Assistance for Entity-Focused
  Tables}. In \bibinfo{booktitle}{\emph{Proceedings of the 40th International
  ACM SIGIR Conference on Research and Development in Information Retrieval}}
  \emph{(\bibinfo{series}{SIGIR '17})}. \bibinfo{pages}{255--264}.
\newblock
\showISBNx{978-1-4503-5022-8}


\bibitem[\protect\citeauthoryear{Zhang and Balog}{Zhang and Balog}{2019}]%
        {Zhang:2019:ADC}
\bibfield{author}{\bibinfo{person}{Shuo Zhang} {and} \bibinfo{person}{Krisztian
  Balog}.} \bibinfo{year}{2019}\natexlab{}.
\newblock \showarticletitle{Auto-completion for Data Cells in Relational
  Tables}. In \bibinfo{booktitle}{\emph{Proceedings of the 28th ACM
  International Conference on Information and Knowledge Management}}
  \emph{(\bibinfo{series}{CIKM '19})}. \bibinfo{pages}{761--770}.
\newblock
\showISBNx{978-1-4503-6976-3}


\bibitem[\protect\citeauthoryear{Zhang and Balog}{Zhang and Balog}{2020}]%
        {Zhang:2020:WTE}
\bibfield{author}{\bibinfo{person}{Shuo Zhang} {and} \bibinfo{person}{Krisztian
  Balog}.} \bibinfo{year}{2020}\natexlab{}.
\newblock \showarticletitle{Web Table Extraction, Retrieval, and Augmentation:
  A Survey}.
\newblock \bibinfo{journal}{\emph{ACM Trans. Intell. Syst. Technol.}}
  \bibinfo{volume}{11}, \bibinfo{number}{2}, Article
  \bibinfo{articleno}{Article 13} (\bibinfo{date}{Jan.} \bibinfo{year}{2020}),
  \bibinfo{numpages}{35}~pages.
\newblock


\bibitem[\protect\citeauthoryear{Zhang, Meij, Balog, and Reinanda}{Zhang
  et~al\mbox{.}}{2020}]%
        {Zhang:NED:2020}
\bibfield{author}{\bibinfo{person}{Shuo Zhang}, \bibinfo{person}{Edgar Meij},
  \bibinfo{person}{Krisztian Balog}, {and} \bibinfo{person}{Ridho Reinanda}.}
  \bibinfo{year}{2020}\natexlab{}.
\newblock \showarticletitle{Novel Entity Discovery from Web Tables}. In
  \bibinfo{booktitle}{\emph{Proceedings of The Web Conference 2020}}
  \emph{(\bibinfo{series}{WWW '20})}. \bibinfo{pages}{1298--1308}.
\newblock


\bibitem[\protect\citeauthoryear{Zhou, Khashabi, Tsai, and Roth}{Zhou
  et~al\mbox{.}}{2018}]%
        {Zhou:2018:ZOE}
\bibfield{author}{\bibinfo{person}{Ben Zhou}, \bibinfo{person}{Daniel
  Khashabi}, \bibinfo{person}{Chen-Tse Tsai}, {and} \bibinfo{person}{Dan
  Roth}.} \bibinfo{year}{2018}\natexlab{}.
\newblock \showarticletitle{Zero-Shot Open Entity Typing as Type-Compatible
  Grounding}. In \bibinfo{booktitle}{\emph{Proceedings of the 2018 Conference
  on Empirical Methods in Natural Language Processing}}.
  \bibinfo{pages}{2065--2076}.
\newblock


\end{thebibliography}
% \balancecolumns

\end{document}